\documentclass[aps,showpacs,twocolumn,unsortedaddress,superscriptaddress]{revtex4-1}
\usepackage{graphicx} 
\usepackage{graphics, bm, psfrag, amsmath, amssymb, epsfig, grffile, float}
\usepackage{color}
\usepackage{hyperref}
\hypersetup{
    colorlinks=true, 
    linktoc=all,     
    linkcolor=blue,  
}

\usepackage{booktabs}
\newcommand{\sgn}{\text{sgn}}
\begin{document}

\title{Excitonic instability in optically-pumped three-dimensional Dirac materials}
\author{Anna Pertsova}
\affiliation{Nordita, Roslagstullsbacken 23, SE-106 91 Stockholm, Sweden}
\author{Alexander V. Balatsky}
\affiliation{Nordita, Roslagstullsbacken 23, SE-106 91 Stockholm, Sweden}
\affiliation{Institute for Materials Science, Los Alamos National Laboratory, Los Alamos New Mexico 87545, USA}
\affiliation{Dept of Physics, University of Connecticut,  Storrs, CT 06269, USA}

\date{today}

\begin{abstract}
Recently it was suggested that transient excitonic instability can be realized in optically-pumped two-dimensional (2D) Dirac materials (DMs), such as 
graphene and topological insulator surface states. 
Here we discuss the possibility of achieving a transient excitonic condensate in optically-pumped three-dimensional (3D) DMs, such as 
Dirac and Weyl semimetals, described by non-equilibrium chemical potentials for photoexcited electrons and holes.
Similar to the equilibrium case with long-range interactions, we find that for pumped 3D DMs with screened Coulomb potential two possible 
excitonic phases exist, an excitonic insulator phase and the charge density wave phase originating from intranodal and internodal interactions, respectively.  
In the pumped case, the critical coupling for excitonic instability vanishes; therefore, the two phases coexist for arbitrarily weak coupling strengths. 
The excitonic gap in the charge density wave phase is always the largest one. 
The competition between screening effects and the increase of the density of states with optical pumping results in a reach phase 
diagram for the transient excitonic condensate. Based on the static theory of screening, we find that under certain conditions for the value of the dimensionless coupling constant
 screening in 3D DMs can be weaker than in 2D DMs. 
Furthermore, we identify the signatures of the transient excitonic condensate that could be probed by scanning tunneling spectroscopy, photoemission and 
optical conductivity measurements. Finally, we provide estimates of critical temperatures and excitonic gaps  
for existing and hypothetical 3D DMs.
\end{abstract}


\maketitle


\section{Introduction}
In the past decade there has been a surge of interest in the so-called Dirac materials (DMs) which exhibit a linear, Dirac-like spectrum of quasiparticle 
excitations~\cite{wehling2014dirac}. 
This rapidly growing class encompasses a diverse range of quantum materials such as high-temperature $d$-wave superconductors~\cite{balatsky_rmp2006}, 
superfluid $^3$He~\cite{volovik_he3}, 
graphene~\cite{neto2009electronic}, topological insulators~\cite{hasan_rmp2010,qi_rmp2011}, and Dirac~\cite{neupane2014dirac} and 
Weyl semimetals~\cite{huang2015weyl,xu2015weyl}. These materials are characterized by the presence of nodes in the quasiparticle spectrum and their properties can be understood 
within a unifying framework of DMs. The concept of DMs has recently been extended to bosonic DMs, e.g. bosonic systems with Dirac nodes in the excitation spectra 
which can be realized, for instance, in 
 various artificial honeycomb lattices~\cite{fransson_prb2016,banerjee_prb2016}. The class of DMs also includes Dirac nodal line semimetals, 
 in which two bands with linear dispersion are degenerate along a one-dimensional curve in momentum space~\cite{burkov_prb2011_linenodes,sun_prb2017_linenodes}.

An important topic that has emerged in the last few years is the study of non-equilibrium dynamics of DMs. One example is the 
interplay between light and the Dirac states in DMs~\cite{Wang_prl2016_light_dm,sanchez-barriga_prx2014_light_dm,Kuroda_prl2016_light_dm}. 
Understanding the non-equilibrium dynamics of Dirac carriers 
subject to perturbations by electromagnetic fields is crucial for applications in ultrafast photonics and high-mobility 
optoelectonics~\cite{otsuji2012graphene,Wang_nanolett2017_DSM_photodetector}. 
Experimental progress in this field is fueled by the availability of time-sensitive probes such as time-resolved pump-probe
 angular-resolved photoemission spectroscopy (ARPES)~\cite{gierz2013graphene,johannsen2013graphene,ulstrup2014bilayer,Johannsen2015Graphene,gierz2015graphene,zhu2015ultrafast,neupane2015gigantic} and 
optical-pump terraherz(THz)-probe spectroscopy~\cite{george2008ultrafast,gilbertson2011tracing,aguilar2015time} that can study the electron dynamics 
on picosecond (ps) and even femtosecond (fs) time-scales. 

Pump-probe experiments on Dirac states in graphene have demonstrated the existence of a broadband 
population inversion~\cite{li2012femtosecond}, a situation when highly excited electrons and holes form two independent Fermi-Dirac distribution 
with separate chemical potentials. This can generate optical gain and is promising for THz lasing applications~\cite{otsuji2012graphene}. 
The lifetime of population inversion in graphene is of the order of $100$ fs~\cite{li2012femtosecond,gierz2013graphene,johannsen2013graphene,ulstrup2014bilayer,Johannsen2015Graphene}. 
Population inversion has also been demonstrated in three-dimensional topological insulators (3D TIs) with much longer lifetimes, ranging 
from few ps ($\tau\approx 3$~ps for Sb$_2$Te$_3$~\cite{zhu2015ultrafast}) to hundreds of ps 
($\tau\approx 400$~ps for bulk-insulating (Sb$_{1-x}$Bi$_x$)$_2$Te$_3$~\cite{sumida2017}). Such long lifetimes are attributed to slow electron-hole recombination. 

Motivated by these experimental results, we recently proposed to search for transient excitonic instability in optically-excited DMs with 
population inversion~\cite{triola_prb2017}. Given the Dirac nature of the spectrum, an inverted population allows the optical tunability of the density of states (DOS) 
of the electrons and holes, effectively offering control of the strength of the Coulomb interaction. The most promising candidate among two-dimensional (2D) materials is 
free-standing graphene pumped by 
circularly polarized light. 3D TIs with specially designed material parameters are also promising due to potentially long lifetimes 
of the optically-excited states. 

In this paper, we focus on transient states in optically pumped 3D DMs such as the newly discovered Dirac and Weyl semimetals. 
These systems exhibit nodes formed by linearly dispersing bands in 3D momentum space. 
In a Dirac semimetal (DSM), the Dirac states are doubly degenerate. The degeneracy can be lifted 
by breaking either time-reversal or spatial-inversion symmetry, resulting in a Weyl semimetal (WSM) in 
which the nodes appear in pairs with opposite chirality. In addition, WSMs display the so-called 
topological Fermi arcs on their surfaces, which connect the bulk projections of the Weyl nodes. 
DSM have been observed in Cd$_3$As$_2$~\cite{liu_natmat2014_dsm} 
and Na$_3$Bi~\cite{liu_science2014_dsm}. WSM have been recently confirmed in TaAs~\cite{xu_science2015_wsm} and signatures 
of the so-called type-II WSM with tilted cones have been seen in WTe$_2$~\cite{feng_prb2016_wsm2}. These materials display a number of 
remarkable properties such as the solid state realization of the chiral anomaly and the resulting negative magnetoresistance 
effect~\cite{Xiong_science_chiral_anomaly,ali_nature2014_weyl_mr}.

So far experimental and theoretical efforts have been focused mostly on equilibrium and steady-state properties of 3D DMs. Characterization of 
 the electronic structure and spin-texture e.g. by ARPES 
 is used to verify the 
 3D Dirac nature of existing and predicted materials. 
 Considerable amount of theoretical work has been done on magnetoelectrical transport~\cite{Tabert_prb2016_transport} and  
 optical properties~\cite{Hosur_prl2012,Ashby_prb2014,Tabet_prb2016_optical} as well as on 
 the role of disorder and interactions~\cite{Wei_prl2012,Wei_prb2014,Wei_prb2014_odd-parity-sc} in DSMs and WSMs. 
 In particular, excitonic instability in equilibrium WSM with chemical potential at the compensation point was studied for the case 
 of short-range~\cite{Wei_prl2012} and long-range~\cite{Wei_prb2014} interactions.
 
Non-equilibrium properties of 3D DMs is a much less explored topic. In contrast to 2D DMs, very little is 
known about non-equilibrium dynamics and relaxation of photoexcited carriers in 3D DMs. However, examples of pump-probe experiments similar to 
those done on graphene and 3D TIs have already appeared in the 
literature~\cite{Manzoni_prl2015_ultrafast_DSM,Ishida_prb2016_relaxation_DSM,Jadidi_THz_pump-probe_SM,Ma_natphys2017_pumping_WSM,Lu_prb2017_ultrafast_DSM}. 
Recent work revealed that ultrafast relaxation of Dirac fermions in Cd$_3$As$_2$ DSM is qualitatively similar to that of graphene~\cite{Lu_prb2017_ultrafast_DSM}. 
Given the growing interest in driven and 
non-equilibrium quantum states of matter and the potential of DMs for high-performance optoelectonic devices~\cite{otsuji2012graphene,Wang_nanolett2017_DSM_photodetector}, 
this trend will continue to gain momentum. 
In this context, we consider the possibility of realizing 
transient many-body states in optically-pumped 3D DMs. We find that external driving combined with the Dirac nature of quasiparticles 
create favorable conditions for transient exitonic instability.

Our theory is based on a low-energy effective model for a Dirac/Weyl system, which includes mean-field interactions and 
screening effects. Metallic screening which is a crucial factor in non-equilibrium, is treated within 
the static random phase approximation. We consider particle-hole instabilities in 3D DMs assuming the existence of non-equilibrium electron and hole populations 
which can be generated by optical pumping (see Fig.~\ref{fig:0}). Similar to the equilibrium case with unscreened interactions~\cite{Wei_prb2014}, 
we find that for screened Coulomb potential two possible excitonic phases exist, an excitonic insulator phase and the charge density wave phase originating 
from intranodal and internodal interactions, respectively. 
The main difference from the equilibrium case is that the critical coupling for excitonic instability vanishes for finite electron and hole chemical potentials. 
Hence, the two phases coexist for arbitrarily weak coupling strengths. However, the excitonic gap produced in the charge density  wave phase is always larger than 
the one in the excitonic insulator phase. 

Contrary to ordinary metals, we find that screening in 3D DMs can be weaker than in 2D DMs depending on the value of the  
the dimensionless coupling constant of the material. 
We present the phase diagrams for the transient excitonic condensate resulting from the competition between screening and the increase of the DOS 
with optical pumping.  We propose several experimental measurements which could probe the existence of the transient excitonic condensate. Finally, 
 we estimate critical temperatures and the size of excitonic gaps for a few realistic cases. 

The paper is organized as follows. In Section~\ref{model} we present the details of our theoretical model. In particular, we discuss the form of the screened 
Coulomb potential in an optically-pumped 3D DM and derive the self-consistent equation for the excitonic gap. In Section~\ref{results} we present the spectroscopic 
features for the transient excitonic states in a pumped 3D DM, namely the spectral function, the density of states, and the optical conductivity. We also discuss  
the dependence of the size of the gap and the critical temperature on the material parameters such as the interaction strength, the chemical potential and 
the Dirac cone degeneracy. Finally, in Section~\ref{concl} we present concluding remarks.
%

  
\section{Theoretical model}\label{model}
\subsection{Hamiltonian of a 3D DM with interactions}\label{Hamiltonian}
\begin{figure*}[ht!]
\centering
\includegraphics[width=0.9\linewidth,clip=tue]{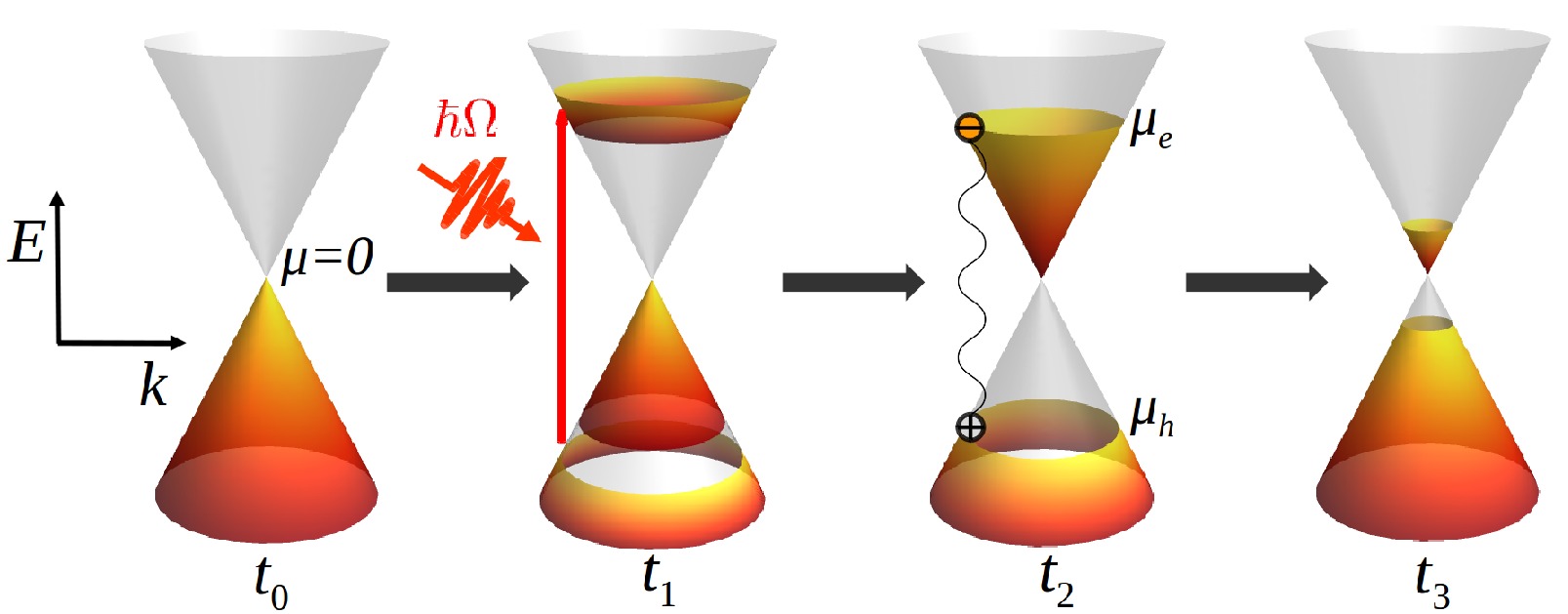}
\caption{Schematic of optically pumped 3D DM with population inversion (for illustration purposes, only a projection of 
the dispersion in 2D momentum space is shown). 
Before the pump, at time $t=t_0$, electrons exist in
equilibrium described by a single chemical potential $\mu=0$; 
at $t=t_1$ electrons are pumped from the valence 
band to the conduction band far from equilibrium; after equilibration time $t=t_2$ 
 electrons and holes can be
described effectively by two distinct Fermi-Dirac distributions
with chemical potentials $\mu_{\mathrm{e}}$ and $\mu_{\mathrm{h}}$; inversion population 
favors pairing between electrons and holes leading to transient excitonic instability;  
transient populations eventually decay towards equilibrium ($t=t_3$).
}
\label{fig:0}
\end{figure*}

A general interacting Hamiltonian for a DSM or WSM can be written as
\begin{equation}
H=H_\mathrm{D/W}+V,\label{eq:gen_Hamil}
\end{equation}
where $H_\mathrm{D/W}$ is the non-interacting Hamiltonian of a DSM/WSM and $V$ contains electron-electron interactions. Below we describe each of the terms in 
Eq.~(\ref{eq:gen_Hamil}).

\subsubsection{Non-interacting Hamiltonian}\label{hamil_nonint}
We define a Weyl node as a linear crossing of two non-degenerate bands in 3D momentum space. A topological number, chirality $\xi=\pm 1$, is assigned to each Weyl node. 
A WSM contains pairs of Weyl nodes with opposite chirality. In a DSM, the two Weyl nodes are degenerate in energy and momentum. 
At low energies, the Hamiltonian $H_\mathrm{D}$ of a DSM can be mapped onto a $4\times 4$ Dirac Hamiltonian, which can be viewed as consisting of two copies of 
a $2\times 2$ Weyl 
Hamiltonian with opposite sign of $\xi$
%
\begin{align}
H_\mathrm{D}&=\sum_{\mathbf{k}}\Psi^{\dagger}_{\mathbf{k}}
\left(
\begin{tabular}{lr}
$H_{+}(\mathbf{k})$ & $0$ \\
$0$ & $H_{-}(\mathbf{k})$
\end{tabular}
\right)\Psi_{\mathbf{k}},\label{eq:H3D_Dirac}\\
H_{\xi}(\mathbf{k})&=\xi\hbar v\mathbf{\sigma}\cdot\mathbf{k}\label{eq:H3D_Weyl}.
\end{align}
%
Here $H_{\xi}$ is the Hamiltonian of a Weyl node with chirality $\xi$, $\mathbf{\sigma}=\{\sigma_x,\sigma_y,\sigma_z\}$ is a set of Pauli matrices, 
$\mathbf{k}=\{k_x,k_y,k_z\}$ is the 3D momentum, 
$v$ is the velocity of the Dirac states and $\Psi_{\mathbf{k}}$ is a four-component spinor.  
Since the Hamiltonian in Eq.~(\ref{eq:H3D_Dirac}) is block-diagonal, in order to describe a DSM, it is sufficient to find the eigenstates of $H_{\xi}$. 
Then the eigenstates of $H_{\mathrm{D}}$ 
are obtained by including a degeneracy factor $g=2$.

In a WSM, the degeneracy between the nodes is lifted  and the Hamiltonian can be written as 
%
%
\begin{align}
H_\mathrm{W}&=\sum_{\xi}\sum_{\mathbf{k}}\Phi^{\dagger}_{\mathbf{k}}H_{\xi}(\mathbf{k})\Phi_{\mathbf{k}},\label{eq:H3D_Weyl0}\\
H_{\xi}(\mathbf{k})&=\xi\hbar v_\mathrm{F}\mathbf{\sigma}\cdot(\mathbf{k}-\xi\mathbf{K})+\xi I K_0\label{eq:HW},
\end{align}
where $\Phi_{\mathbf{k}}$ 
is a two-component spinor and 
$I$ is a $2\times{2}$ identity matrix. Finite $\mathbf{K}$ corresponds to broken time reversal symmetry while finite $K_0$ corresponds to 
broken inversion symmetry. For $\mathbf{K}\ne 0$ and $K_0=0$, the Weyl nodes are located at the same energy but are shifted in momentum space by $2\mathbf{K}$. 
For $\mathbf{K}=0$ and $K_0\ne 0$, the Weyl nodes are located at the same momentum but are shifted in energy by $2K_0$.
As a limiting case at $\mathbf{K}=0$ and $K_0=0$, the nodes become degenerate in energy and momentum as in a DSM [Eq.~(\ref{eq:H3D_Dirac})]. 

Here we focus on the time reversal broken case. Inversion symmetry breaking can be included simply by introducing a rigid shift $\pm K_0$ to the energy eigenvalues. 
To simplify notations, we assign labels R(right) and L(left) for the node located at 
 $\mathbf{q}=\mathbf{k}-\mathbf{K}$ with chirality $\xi=+1$ and $\mathbf{q}=\mathbf{k}+\mathbf{K}$ with chirality $\xi=-1$, respectively. The resulting Hamiltonian for 
 R/L node reads
\begin{equation}
H_{R/L}(\mathbf{q})=\pm\hbar v\mathbf{\sigma}\cdot\mathbf{q}.\label{eq:HW_LR}
\end{equation}
The eigenvalues of this Hamiltonian are given by 
$\varepsilon_{\mathbf{q}}^{\pm}=\pm\hbar{v}|\mathbf{q}|$, where $+$ ($-$) stands for conduction (valence) band. The corresponding normalized eigenvectors 
for R/L node are given by 
\begin{eqnarray}
\chi_{\mathbf{q},+}^{R}=
\left(
\begin{array}{l}
 \cos{\frac{\theta}{2}}e^{-i\phi} \\ 
 \sin{\frac{\theta}{2}}
\end{array}
\right); 
\chi_{\mathbf{q},-}^{R}=
\left(
\begin{array}{l}
 -\sin{\frac{\theta}{2}}e^{-i\phi}\label{eq:Weyl_eigenv1}\\ 
 \cos{\frac{\theta}{2}}
\end{array}
\right)\\
\chi_{\mathbf{q},+}^{L}=
\left(
\begin{array}{l}
 -\sin{\frac{\theta}{2}}e^{-i\phi} \\ 
 \cos{\frac{\theta}{2}}
\end{array}
\right); 
\chi_{\mathbf{q},-}^{L}=
\left(
\begin{array}{l}
 \cos{\frac{\theta}{2}}e^{-i\phi}\label{eq:Weyl_eigenv2}\\ 
 \sin{\frac{\theta}{2}}
\end{array}
\right).
\end{eqnarray}

The Hamiltonian $H_{R/L}(\mathbf{q})$ is diagonalized by a unitary transformation ${U^{R/L}}^{\dagger}H_{R/L}(\mathbf{q})U^{R/L}$, where $U^{R/L}$ is composed of 
the eigenvectors in Eqs.~(\ref{eq:Weyl_eigenv1}-\ref{eq:Weyl_eigenv2}) [see Appendix~\ref{appendixA} for details]. 
The spinors $\Phi_{\mathbf{q}}^{R/L}$ can be expressed in the diagonal basis as 
 $\Phi_{\mathbf{q},\sigma}^{R/L}=\sum_{n=\pm}\Phi_{\mathbf{q},n,\sigma}^{R/L}$, where 
 $\Phi_{\mathbf{q},n,\sigma}^{R/L}=\chi_{\mathbf{q},n}^{R/L,\sigma} c_{\mathbf{q},n}^{R/L}$ and ${c_{\mathbf{q},n}^{R/L}}^{\dagger}(c_{\mathbf{q},n}^{R/L})$, $n=\pm$ are the 
 fermionic creation(annihilation) operators corresponding to the bands $\varepsilon_{\mathbf{q}}^{n}$. 
 
\subsubsection{Coulomb interactions in Dirac/Weyl Semimetal}\label{hamil_int}
We will now consider electron-electron interactions for a system of two, in general non-degenerate, Weyl nodes. The interacting Hamiltonian for a DSM can then be obtained 
 in the limiting case when the nodes are degenerate in momentum space and energy. Starting from a general spin-independent particle-particle interaction potential 
\begin{equation}
 V=\sum_{\sigma,\sigma^{\prime}}\sum_{\mathbf{k},\mathbf{k}',\mathbf{q}}\sum_{\alpha_i=R,L}{\Phi_{\mathbf{k}'+\mathbf{q},\sigma'}^{\alpha_1{\dagger}}}
 \Phi_{\mathbf{k}',\sigma'}^{\alpha_2} {\Phi_{\mathbf{k}-\mathbf{q},\sigma}^{\alpha_3{\dagger}}}
 \Phi_{\mathbf{k},\sigma}^{\alpha_4},
\end{equation}
we express the interaction in the diagonal basis of fermionic operators $c_{\mathbf{q},n}^{R/L}$. Our derivation follows Refs.~\cite{Wei_prl2012,Wei_prb2014}.  
 For completeness, the main steps of the derivation are summarized in Appendix~\ref{appendixB}. We change the notations 
 as $\mathbf{k}\rightarrow\mathbf{q}$, where $q=\mathbf{k\mp\mathbf{K}}$ for R/L node and express the wavefunctions as   
$\Phi_{\mathbf{q},\sigma}^{R/L}=\sum_{n=\pm}\Phi_{\mathbf{q,n,\sigma}}^{R/L}=\chi_{\mathbf{q},n}^{R/L,\sigma} c_{\mathbf{q},n}^{R/L}$. 
We also use the fact that $V(\mathbf{k})=V(-\mathbf{k})$. Considering pairing of the form 
$\Phi_{\mathbf{q},n,\sigma}^{\alpha\dagger}\Phi_{\mathbf{q},-n,\sigma'}^{\beta}$, $\alpha,\beta=\mathrm{R/L}$, the interaction potential 
is given by 
%
%
%
\begin{widetext}
\begin{eqnarray}
V=-\sum_{\substack{\mathbf{q},\mathbf{q}' \\ n=\pm}}\left\lbrace\right.
 V(\mathbf{q}-\mathbf{q}')A(\mathbf{q},\mathbf{q}')\sum_{\alpha=R,L}
 c_{\mathbf{q},n}^{\alpha\dagger}c_{\mathbf{q},-n}^{\alpha}c_{\mathbf{q}',-n}^{\alpha\dagger}c_{\mathbf{q}',n}^{\alpha}
 +V(\mathbf{q}-\mathbf{q}'-2\mathbf{K})B(\mathbf{q},\mathbf{q}')
 c_{\mathbf{q},n}^{L\dagger}c_{\mathbf{q},-n}^{L}c_{\mathbf{q}',-n}^{R\dagger}c_{\mathbf{q}',n}^{R}\nonumber\\
 -\left[2V(2\mathbf{K})-V(\mathbf{q}-\mathbf{q}')C(\mathbf{q},\mathbf{q}')\right]
  c_{\mathbf{q},n}^{L\dagger}c_{\mathbf{q},-n}^{R}c_{\mathbf{q}',-n}^{R\dagger}c_{\mathbf{q}',n}^{L}\left.\right\rbrace
  \label{eq:V_Coulomb},
\end{eqnarray}
\end{widetext}
where $A,B,C$ are momentum-dependent coefficients, defined as
\begin{eqnarray}
 A(\mathbf{q},\mathbf{q}')&=&\frac{\hat{e}_{q}\cdot\hat{e}_{q'}^{*}+\hat{e}_q^{*}\cdot \hat{e}_{q'}}{4}\nonumber\\
                          &=&\frac{\sin\theta\sin\theta'}{2}+\frac{1+\cos\theta\cos\theta'}{2}\cos(\phi-\phi'),\nonumber\\
\end{eqnarray}

\begin{eqnarray}
 B(\mathbf{q},\mathbf{q}')&=&\frac{\hat{e}_{q}\cdot\hat{e}_{q'}+\hat{e}_q^{*}\cdot \hat{e}_{q'}^{*}}{2}\nonumber\\
                          &=&\frac{\sin\theta\sin\theta'}{2}-\frac{1-\cos\theta\cos\theta'}{2}\cos(\phi-\phi'),\nonumber\\ 
\end{eqnarray}

\begin{eqnarray}
 C(\mathbf{q},\mathbf{q}')&=&\hat{q}\cdot\hat{q}'+1.
\end{eqnarray}

In the above expressions, $\hat{e}_q=\hat{e}_q^1+i\hat{e}_{q}^2$ and 
$\left\lbrace \hat{q},\hat{e}_q^1,\hat{e}_q^2\right\rbrace\equiv\left\lbrace \hat{r},\hat{\theta},\hat{\phi}\right\rbrace$, 
where $\theta$ and $\phi$ are the azimuthal and polar angles of the spherical coordinate system respectively. 
In Eq.~(\ref{eq:V_Coulomb}), $2\mathbf{K}$ is the distance in momentum space between the two Weyl nodes. 
DSM is obtained by setting $\mathbf{K}=0$. In Eq.~(\ref{eq:V_Coulomb}), the first two terms correspond to intranodal interactions (pairing 
within a single node, L or R) 
while the last term corresponds to internodal 
interactions (pairing between L and R nodes). 

Two forms of the interaction potential $V(\mathbf{q})$ can be considered: 
(i) a simplified short-range, or contact interaction, i.e.
$V(\mathbf{q})=V_0$, where $V_0$ is a constant in momentum space (delta function in real space), (ii) a more realistic Coulomb potential. 
In equilibrium, i.e. when the chemical potential is exactly at the Dirac or Weyl node, 
one can in principle consider the long-range unscreened Coulomb potential, $V(\mathbf{q})\propto \frac{1}{|\mathbf{q}|^2}$ ~\cite{Wei_prb2014}. 
However, for chemical potential away from the node, which is the case for doping 
and optical pumping, screening is important. 
In the following section we will derive the expressions for the screened Coulomb potential within 
the static random phase approximation for both 2D and 3D DMs. 

\subsubsection{Screened Coulomb potential in 2D and 3D DM}\label{TF_screening}
The general expression for the frequency-dependent dielectric function $\varepsilon({\omega,\mathbf{q}})$ in the random phase approximation, or the Lindhard dielectric function, 
reads~\cite{haug2004quantum}
\begin{equation}
\varepsilon(\textbf{q},\omega)=1-V({\mathbf{q}})\sum_{\mathbf{k}}\frac{f_{\mathbf{k}-\mathbf{q}}-f_\mathbf{k}}{\omega+i\delta+\omega_{\mathbf{k-q}}-\omega_{\mathbf{k}}},
\label{eq:Lindhard_dyn}
\end{equation} 
where $V(\mathrm{q})$ is the Coulomb potential, $f_{\mathbf{k}}$ are the Fermi factors, $\omega_{\mathbf{k}}=\varepsilon_{\mathbf{k}}/\hbar$ and 
$\varepsilon_\mathbf{k}$ are the energies. In the limit $\omega\rightarrow 0$, 
we obtain the static dielectric function $\varepsilon(\mathbf{q},0)$, the statically screened Coulomb potential $V_s(\mathbf{q})$,  
and the expression for the screening wavevector $\kappa$. In the 2D case, we have
\begin{align}
\varepsilon^{\mathrm{2D}}(\mathbf{q},0)&=1+\frac{\kappa_{\mathrm{2D}}}{q},\label{eq:Lindhart_stat_2D}\\
V_s^{\mathrm{2D}}(\mathbf{q})&=\frac{V^{\mathrm{2D}}({\mathbf{q}})}{\varepsilon^{\mathrm{2D}}(\mathbf{q},0)}=\frac{2\pi e^2}{\varepsilon}\frac{1}{q+\kappa_{\mathrm{2D}}},\label{eq:coul_scr_2D}\\
\kappa_{\mathrm{2D}}&=\frac{2\pi e^2}{\varepsilon}\frac{\partial n}{\partial \mu},\label{eq:scr_wv_2D}
\end{align}
where $n$ is the electron density, $\mu$ is the chemical potential, $e$ is the electron charge and $\varepsilon$ is the dielectric constant. Here 
$V^{\mathrm{2D}}({\mathbf{q}})=\frac{2\pi e^2}{\varepsilon}\frac{1}{q}$ is the 
unscreened Coulomb potential in 2D momentum space, which is obtained by Fourier transform from the bare real space potential $V(r)=\frac{e^2}{\varepsilon r}$. 

Analogously, in the 3D case, we have
\begin{align}
\varepsilon^{\mathrm{3D}}(\mathbf{q},0)&=1+\frac{{\kappa_{\mathrm{3D}}}^2}{q^2},\label{eq:Lindhart_stat_3D}\\
V_s^{\mathrm{3D}}({\mathbf{q}})&=\frac{V^{\mathrm{3D}}({\mathbf{q}})}{\varepsilon^{\mathrm{3D}}(\mathbf{q},0)}=\frac{4\pi e^2}{\varepsilon}\frac{1}{q^2+{\kappa_{\mathrm{3D}}}^2}\,
\label{eq:coul_scr_3D}\\
\kappa_{\mathrm{3D}}&=\sqrt{\frac{4\pi e^2}{\varepsilon}\frac{\partial n}{\partial \mu}},\label{eq:scr_wv_3D}
\end{align}
where $V^{\mathrm{3D}}({\mathbf{q}})=\frac{4\pi e^2}{\varepsilon}\frac{1}{q^2}$ is the 
unscreened Coulomb potential in 3D momentum space.

So far we assumed the presence of one type of carriers, say electrons, defined by density $n$ and chemical potential $\mu$. 
In the case of population inversion generated by optical pumping, we have electron and hole plasmas which exist 
at different densities and chemical potentials. Therefore, one should define the global screening wavevector~\cite{klingshirn2005optics}
\begin{align}
\kappa_\mathrm{2D}&=\frac{2\pi e^2}{\varepsilon}\sum_{i=e,h}\frac{\partial n_i}{\partial \mu_i}\label{eq:tot_scr_wv_2D},\\
\kappa_\mathrm{3D}&=\sqrt{\frac{4\pi e^2}{\varepsilon}\sum_{i=e,h}\frac{\partial n_i}{\partial \mu_i}}\label{eq:tot_scr_wv_3D}.
\end{align}
where $n_{i}$ and  $\mu_i$ ($i=e,h$) are the electron/hole density and electron/hole chemical potential, respectively. 
It is instructive to re-write the global screening wavevector in terms of the screening vectors of electron and hole plasmas 
\begin{align}
\kappa_\mathrm{2D}&=\kappa_\mathrm{2D}^{e}+\kappa_\mathrm{2D}^{h}\label{eq:tot_scr_wv_2D_2},\\
(\kappa_\mathrm{3D})^2&=(\kappa_\mathrm{3D}^{e})^2+(\kappa_\mathrm{3D}^{h})^2\label{eq:tot_scr_wv_3D_2},
\end{align}
where $\kappa_\mathrm{2D}^{e/h}$ and $\kappa_\mathrm{2D}^{e/h}$ are given in Eq.~(\ref{eq:scr_wv_2D}) and (\ref{eq:scr_wv_3D}), respectively. 
Assuming equal densities for electrons and holes, one can see that in 2D the screening wavevector increases by a factor of $2$ while in 
3D it increases by a factor of $\sqrt{2}$, compared to electron/hole screening wavevector. 

Using the general expressions obtained above, we calculated the screening wavevector in 2D and 3D for a system with Dirac dispersion. 
The results are summarized in Table~\ref{Table:screening} (the details of the calculation are presented in Appendix~\ref{appendixC}, where we
 also show the results for 2D and 3D electron gas). 
Only the results for a single type of carriers (electrons or holes) are presented. The total screening can be then 
 obtained from Eqs.~(\ref{eq:tot_scr_wv_2D_2}) and (\ref{eq:tot_scr_wv_3D_2}). We take the zero temperature limit, $T\rightarrow 0$, which is referred to 
 as the Thomas-Fermi approximation~\cite{haug2004quantum,dassarma2011graphene,dassarma2014_3dDirac}.

\begin{table}[ht!]
\caption{Density of states $D(E)$ as a function of energy and the Thomas-Fermi screening wavevector $\kappa_\mathrm{TF}$ 
for a Dirac spectrum. 
$\alpha$ is the dimensionless coupling constant of a DM, $k_\mathrm{F}$ is the Fermi wavevector (see text for definitions), $g$ is the 
Dirac cone degeneracy. 
}
\begin{tabular}{r||c|c}
System  & $D(E)$  & $\kappa_\mathrm{TF}$ \\
\hline
\hline
2D DM &  $g E/2\pi(\hbar v)^2$ & $g\alpha k_\mathrm{F}$\\
\hline
3D DM &  $g E^2/2\pi^2(\hbar v)^3$ & $\sqrt{2 g\alpha/\pi}k_\mathrm{F}$ \\
\end{tabular}
\label{Table:screening}
\end{table}

In Table~\ref{Table:screening}, we defined the dimensionless 
coupling constant $\alpha=e^2/\hbar\varepsilon v$ and the Fermi wavevector $k_\mathrm{F}=\mu/\hbar{v}$. One can see that in both 2D and 3D DM, 
the screening wavevector scales linearly with $k_\mathrm{F}$. However, the prefactors in the linear dependence are different. 
The dimensionless coupling constant 
$\alpha$ and the degeneracy factor $g$ can be different for 2D and 3D DM. 
One needs to know the values for these parameters in order to make quantitative predictions 
about the strength of the screening effects. Assuming for simplicity equal chemical potentials and velocities of the Dirac states in 2D and 3D DM, 
we find that, in general, the screening in 2D DM is stronger than 
in 3D DM if the following conditions are satisfied
\begin{eqnarray}
 \alpha_\mathrm{3D} & < \pi g_\mathrm{2D}^2 \alpha_\mathrm{2D}^2/2 g_\mathrm{3D},\quad\mathrm{for}\quad g_\mathrm{2D} \alpha_\mathrm{2D}>1,\\
 \alpha_\mathrm{3D} & > \pi g_\mathrm{2D}^2 \alpha_\mathrm{2D}^2/2 g_\mathrm{3D},\quad\mathrm{for}\quad g_\mathrm{2D} \alpha_\mathrm{2D}<1,
\end{eqnarray}
where the subscript $\mathrm{2D/3D}$ refers to 2D/3D DM.

\subsection{Excitonic instability in a pumped 3D DM: quasi-equilibrium model}\label{Exc_Quasi_equil}
Before discussing excitonic instability in pumped 3D DM, we should note that conditions for excitonic condensation can be realized without pumping in WSM with broken spatial inversion symmetry, 
when the nodes are shifted symmetrically in energy with respect to the original Dirac node. In this case, there exist perfectly nested electron and hole Fermi surfaces, similarly 
to the case of graphene in parallel magnetic field~\cite{aleiner_prb2007}. Hence the excitonic order can be established at arbitrary weak coupling strength. 
Such situation has been considered for instance in Ref.~\cite{zyuzin_prb2012_ex_wsm}. The case of optical pumping considered here is unique in a sense that the excitonic states 
are of transient nature. In contract to the equilibrium case, the electron and hole DOS, and hence the strength of the effective interaction and the value of critical temperature, 
is controlled by optical pumping~\cite{triola_prb2017}.  

We will now consider a simple model of an optically pumped DM, in which electrons in conduction and valence bands are described by 
two separate Fermi-Dirac distributions with different chemical potentials $\mu_e$ and $\mu_h$, respectively, as illustrated in Fig.~\ref{fig:0}. 
We assume that these non-equilibrium populations have been established 
after some time $t_2$ and we will solve for the excitonic order parameter self-consistently assuming quasi-equilibrium, i.e. that the lifetime $\tau$ of the inverted population is infinitely long. 
The relaxation of the populations and the order parameter towards equilibrium using a dynamical method based on rate equations~\cite{triola_prb2017} will be considered in future work.

We consider the Hamiltonian for a system of two Weyl nodes, with interactions given by Eq.~(\ref{eq:V_Coulomb}). For the case of optical pumping, 
the band dispersions of conduction electrons and valence electrons (holes) need to be modified as follows
\begin{eqnarray}
 \varepsilon^{+}_{\mathbf{q}}&=&\hbar{v}|\mathbf{q}|-\mu_e\label{bands_pumped_CB}\\
 \varepsilon^{-}_{\mathbf{q}}&=&-\hbar{v}|\mathbf{q}|-\mu_h\label{bands_pumped_VB}.
\end{eqnarray}

The self-consistent equation for the order parameter, or gap $\Delta(\mathbf{q})$, of the excitonic condensate in this system reads (see Appendix~\ref{appendixD} for the derivation)
\begin{equation}\label{gap_pumped}
 \Delta(\mathbf{q})=\sum_{\mathbf{q}'}\tilde{V}(\mathbf{q},\mathbf{q}')\frac{\Delta(\mathbf{q}')}{\omega_{+}({\mathbf{q}'})-\omega_{-}({\mathbf{q}'})}
 [n_{\mathrm{F}}(\omega_{+})-n_{\mathrm{F}}(\omega_{-})],
\end{equation}
where 
\begin{equation}
 \omega_{\pm}(\mathbf{q})=\frac{\varepsilon^{+}_{\mathbf{q}}+\varepsilon^{-}_{\mathbf{q}}}{2}\pm\frac{1}{2}
 \sqrt{(\varepsilon^{+}_{\mathbf{q}}-\varepsilon^{-}_{\mathbf{q}})^2+4|\Delta(\mathbf{q})|^2}\label{exc_bands}
\end{equation}
are the excitonic bands, $n_{\mathrm{F}}(\omega)=1/(e^{\omega/k_{\mathrm{B}}T}+1)$ is the Fermi-Dirac distribution, $T$ is the temperature (assumed to be 
the same for photoexcited electrons and holes), and $k_{\mathbf{B}}$ is the Boltzmann constant. 
The form of the interaction potential $\tilde{V}(\mathbf{q},\mathbf{q}')$ depends on the particular case we are considering 
(intranodal or internodal scattering) and the approximations made to the Coulomb potential. 
The order parameter is of the form
\begin{equation}
 \Delta(\mathbf{q})=\sum_{\mathbf{q}'}\tilde{V}(\mathbf{q},\mathbf{q}')\left\langle c^{\alpha\dagger}_{\mathbf{q}',n} c^{\beta}_{\mathbf{q}',-n} \right\rangle, 
\end{equation}
where $\alpha,\beta=L,R$. The case $\alpha=\beta$ corresponds to intranodal interactions and leads to an excitonic insulator (EI) phase. The case $\alpha\ne \beta$ corresponds to internodal 
interactions and leads to 
a charge density wave (CDW) phase with the modulation momentum equal to the distance $2\mathbf{K}$ between the Weyl nodes. 
Excitonic phases in a WSM for contact and unscreened Coulomb potential in equilibrium ($\mu_e=-\mu_h=0$) have been studied in detail in Refs.~\cite{Wei_prl2012,Wei_prb2014}. 
It was shown that for contact (short-range) interaction, the EI phase is more energetically favorable and is accompanied by gap opening at the nodes. In contract, for long-range 
unscreened interactions the CDW phase is more energetically favorable. Below we specify the form of the order parameter and the gap equation for intranodal and internodal interactions 
in the case of a 
pumped (non-equilibrium) 3D DM. 
(The equilibrium case is considered in Appendix~\ref{appendixD} as a limit 
of Eq.~(\ref{gap_pumped}) for $\mu_e=-\mu_h=0$.) 

\subsubsection{Excitonic Insulator (intranodal interactions)}\label{Exc_equil_EI}
The intranodal part of the interaction potential is given by the first two terms in Eq.~(\ref{eq:V_Coulomb}). 
Since $|\mathbf{q}-\mathbf{q}'|<<|\mathbf{q}-\mathbf{q}'-2\mathbf{K}|$, we can neglect the second term. The interaction potential becomes
\begin{equation}
 V=-\sum_{\mathbf{q},\mathbf{q}',n=\pm}\tilde{V}_\mathrm{intra}(\mathbf{q},\mathbf{q}')\sum_{\alpha=L,R}c_{\mathbf{q},n}^{\alpha\dagger}c_{\mathbf{q},-n}^{\alpha}
 c_{\mathbf{q}',-n}^{\alpha\dagger}c_{\mathbf{q}',n}^{\alpha},
\end{equation}
where
\begin{eqnarray}\label{eq:V_intra}
\tilde{V}_\mathrm{intra}(\mathbf{q},\mathbf{q}')&=&V(\mathbf{q}-\mathbf{q}')[\frac{\sin\theta\sin\theta'}{2}+\frac{1+\cos\theta\cos\theta'}{2}\nonumber\\
&\phantom{=}&\phantom{V(\mathbf{q}-\mathbf{q}')[\frac{\sin\theta\sin\theta'}{2}}\times\cos(\phi-\phi')].
\end{eqnarray}
Furthermore, we only keep the slowest varying term in the angular-dependent part of the potential given by $\cos(\phi-\phi')/2$. 
The mean-field order parameter can be defined as
\begin{equation}
 \Delta^{\alpha}(\mathbf{q})=\sum_{\mathbf{q}'}\tilde{V}_\mathrm{intra}(\mathbf{q},\mathbf{q}')\left\langle c_{\mathbf{q}',+}^{\alpha\dagger}c_{\mathbf{q},-}^{\alpha}\right\rangle,\quad\alpha=L,R.
\end{equation}
The mean-field Hamiltonian of the system reads
\begin{eqnarray}
 H&=&\sum_{\substack{\mathbf{q},\alpha=L/R \\ n=\pm}}\varepsilon_{\mathbf{q}}^{n}c_{\mathbf{q},n}^{\alpha\dagger}c_{\mathbf{q},n}^{\alpha}
 -\sum_{\mathbf{q},\alpha=L,R}\tilde{\Delta}^{\alpha}(\mathbf{q})c_{\mathbf{q},-}^{\alpha\dagger}c_{\mathbf{q},+}^{\alpha}\nonumber\\
 &-&\sum_{\mathbf{q},\alpha=L,R}\tilde{\Delta}^{\alpha*}(\mathbf{q})c_{\mathbf{q},+}^{\alpha\dagger}c_{\mathbf{q},-}^{\alpha},\label{eq:EI_Hamil}
\end{eqnarray}
where $\tilde{\Delta}^{\alpha}(\mathbf{q})=2\Delta^{\alpha}(\mathbf{q})$ and $\varepsilon^n_{\mathbf{q}}$ for $n=\pm$ are given 
in Eqs.~(\ref{bands_pumped_CB})-(\ref{bands_pumped_VB}). The first term in Eq.~(\ref{eq:EI_Hamil}) is the non-interacting 
Hamiltonian of the two nodes while the last two terms describes interactions within each node. 

The self-consistent gap equation becomes
\begin{equation}
 \Delta^{\alpha}(\mathbf{q})=
\sum_{\mathbf{q}'}2\tilde{V}_\mathrm{intra}\frac{\Delta(\mathbf{q}')}{\omega_{+}({\mathbf{q}'})-\omega_{-}({\mathbf{q}'})}
 [n_{\mathrm{F}}(\omega_{+})-n_{\mathrm{F}}(\omega_{-})],
\end{equation}
where $\omega_{\pm}$ are defined in Eq.~(\ref{exc_bands}).

\subsubsection{Charge density wave (internodal interactions)}\label{Exc_equil_CDW}
The internodal part of the interaction potential is given by the last term in Eq.~(\ref{eq:V_Coulomb}). 
Since $|\mathbf{q}-\mathbf{q}'|<<|\mathbf{K}|$, the leading term is is given by $V(\mathbf{q}-\mathbf{q}')(1+\hat{q}\cdot\hat{q}')$
 and the interaction potential can be written as
\begin{equation}
 V=-\sum_{\mathbf{q},\mathbf{q}',n=\pm}\tilde{V}_\mathrm{inter}(\mathbf{q},\mathbf{q}')c_{\mathbf{q},n}^{L\dagger}c_{\mathbf{q},-n}^{R}c_{\mathbf{q}',-n}^{R\dagger}c_{\mathbf{q}',n}^{L},
\end{equation}
where
\begin{equation}\label{eq:V_inter}
\tilde{V}_\mathrm{inter}(\mathbf{q},\mathbf{q}')=-\left[2V(2\mathbf{K})-V(\mathbf{q}-\mathbf{q}')(1+\hat{q}\cdot\hat{q}')\right].
\end{equation}
Furthermore, we neglect the angular-dependent part proportional to $\hat{q}\cdot\hat{q}'$. In this case, the mean-field order parameter is defined as 
\begin{equation}
 \Delta_n(\mathbf{q})=\sum_{\mathbf{q}'}\tilde{V}_{inter}(\mathbf{q},\mathbf{q}')\left\langle c_{\mathbf{q}',-n}^{R\dagger}c_{\mathbf{q},n}^{L}\right\rangle,
\end{equation}
and the mean-field Hamiltonian is given by
\begin{eqnarray}
 H&=&\sum_{\substack{\mathbf{q},\alpha=L/R \\ n=\pm}}\varepsilon_{\mathbf{q}}^{n}c_{\mathbf{q},n}^{\alpha\dagger}c_{\mathbf{q},n}^{\alpha}-
 \sum_{\mathbf{q},n=\pm}\Delta_n(\mathbf{q})c_{\mathbf{q},n}^{L\dagger}c_{\mathbf{q},-n}^{R}\nonumber\\
 &-&\sum_{\mathbf{q},n=\pm}\Delta_n^{*}(\mathbf{q})c_{\mathbf{q},-n}^{R\dagger}c_{\mathbf{q},n}^{L}.\label{eq:CDW_Hamil}
\end{eqnarray}

The corresponding self-consistent gap equation reads
\begin{equation}
 \Delta_n(\mathbf{q})=\sum_{\mathbf{q}'}\tilde{V}_\mathrm{inter}(\mathbf{q},\mathbf{q}')\frac{\Delta(\mathbf{q}')}{\omega_{+}({\mathbf{q}'})-\omega_{-}({\mathbf{q}'})}.
\end{equation}

\subsubsection{Numerical solution of the gap equation}\label{Exc_equil_numer}
Since the order parameter is momentum-dependent, we will solve the self-consistent gap equation numerically. 
In order to simplify the numerical analysis, we will use dimensionless units  $\Delta\rightarrow\Delta/\hbar{v}\lambda$,
  $\mathbf{q}\rightarrow\mathbf{q}/\lambda$, 
  $T\rightarrow{k_\mathrm{B}T}/{\hbar{v}\lambda}$, 
%
%
where $\lambda$ is the momentum cutoff of the Dirac model in Eqs.~(\ref{bands_pumped_CB})-(\ref{bands_pumped_VB}). The corresponding cutoff energy scale is $\Lambda\equiv\hbar{v}\lambda$.

In what follows we will assume that $V(\mathbf{q}-\mathbf{q}')$ is given by the 3D screened Coulomb potential 
\begin{eqnarray}
  V(\mathbf{q}-\mathbf{q}')&=&\frac{4\pi}{\varepsilon}\frac{1}{(\mathbf{q}-\mathbf{q}')^2+\kappa^2}\nonumber\\
                           &\equiv&\frac{4\pi\alpha\hbar{v}}{(\mathbf{q}-\mathbf{q}')^2+\kappa^2},
\end{eqnarray}
where $\kappa\equiv\kappa_\mathrm{3D}$ is a 3D Thomas-Fermi screening wavevector (see Section~\ref{TF_screening}). Hence, taking into account only the leading terms in the 
interaction potential,  
the gap equation in dimensionless units for intranodal an internodal interactions is given by respectively

\begin{eqnarray}\label{eq:EI_gap_eq_final}
  \Delta^{\alpha}(\mathbf{q})&=\frac{4\pi{\alpha}}{(2\pi)^3}\int\frac{1}{(\mathbf{q}-\mathbf{q}')^2+\kappa^2}\frac{\Delta^{\alpha}(\mathbf{q}')\cos(\phi-\phi')}
  {\omega_{+}({\mathbf{q}'})-\omega_{-}({\mathbf{q}'})}\nonumber\\
  &\phantom{=}\phantom{\frac{4\pi{\alpha}}{(2\pi)^3}\int\frac{1}{(\mathbf{q}-\mathbf{q}')^2+\kappa^2}}\times[n_{\mathrm{F}}(\omega_{+})-n_{\mathrm{F}}(\omega_{-})]dV,
\end{eqnarray}
\begin{eqnarray}\label{eq:CDW_eq_final}
  \Delta_n(\mathbf{q})&=\frac{4\pi{\alpha}}{(2\pi)^3}\int\frac{1}{(\mathbf{q}-\mathbf{q}')^2+\kappa^2}\frac{\Delta_n(\mathbf{q}')}{\omega_{+}({\mathbf{q}'})-\omega_{-}({\mathbf{q}'})}\nonumber\\
  &\phantom{=}\phantom{\frac{4\pi{\alpha}}{(2\pi)^3}\int\frac{1}{(\mathbf{q}-\mathbf{q}')^2+\kappa^2}}\times[n_{\mathrm{F}}(\omega_{+})-n_{\mathrm{F}}(\omega_{-})]dV,
\end{eqnarray}

%
where $dV=|\mathbf{q}'|^2 d{q}'\sin\theta'{d}\theta'{d}\phi'$ is the volume element. 

First, we consider the gap equation for the CDW phase, Eq.~(\ref{eq:CDW_eq_final}). In this case, the order parameter is isotropic and the Coulomb potential can be 
replaced by its angle-average, which only depends on the lengths of vectors $\mathbf{q}$ and $\mathbf{q}'$ and 
can be calculated analytically (see Appendix~\ref{appendixD} for details).

The EI gap equation, Eq.~(\ref{eq:EI_gap_eq_final}), is explicitly angle-dependent. However, it can be re-written in the form similar to the CDW case. For this we 
express $\cos(\phi-\phi')$ as $(e^{i(\phi-\phi')}+e^{i(\phi'-\phi}))/2$ and the EI gap as $\Delta^{\alpha}(\mathbf{q})=\Delta^{\alpha}(q)\cdot e^{i\phi}$. The resulting self-consistent 
equation for the magnitude of the gap  $\Delta^{\alpha}(q)$ is identical to the one in the CDW case with additional factor of $1/2$. Therefore the EI gap is always smaller than the 
 CDW gap for the same model parameters.
 
Using the self-consistently calculated order parameter, we can obtain the observable quantities, such as the DOS and the spectral function as well as the phase diagrams for the 
excitonic condensate in pumped 3D DM. The results of the calculations are presented in the next section. 

\section{Results}\label{results}
\subsection{Density of states and spectral function}\label{DOS_Spec}
\begin{figure}[ht!]
\includegraphics[width=0.98\linewidth,clip=true]{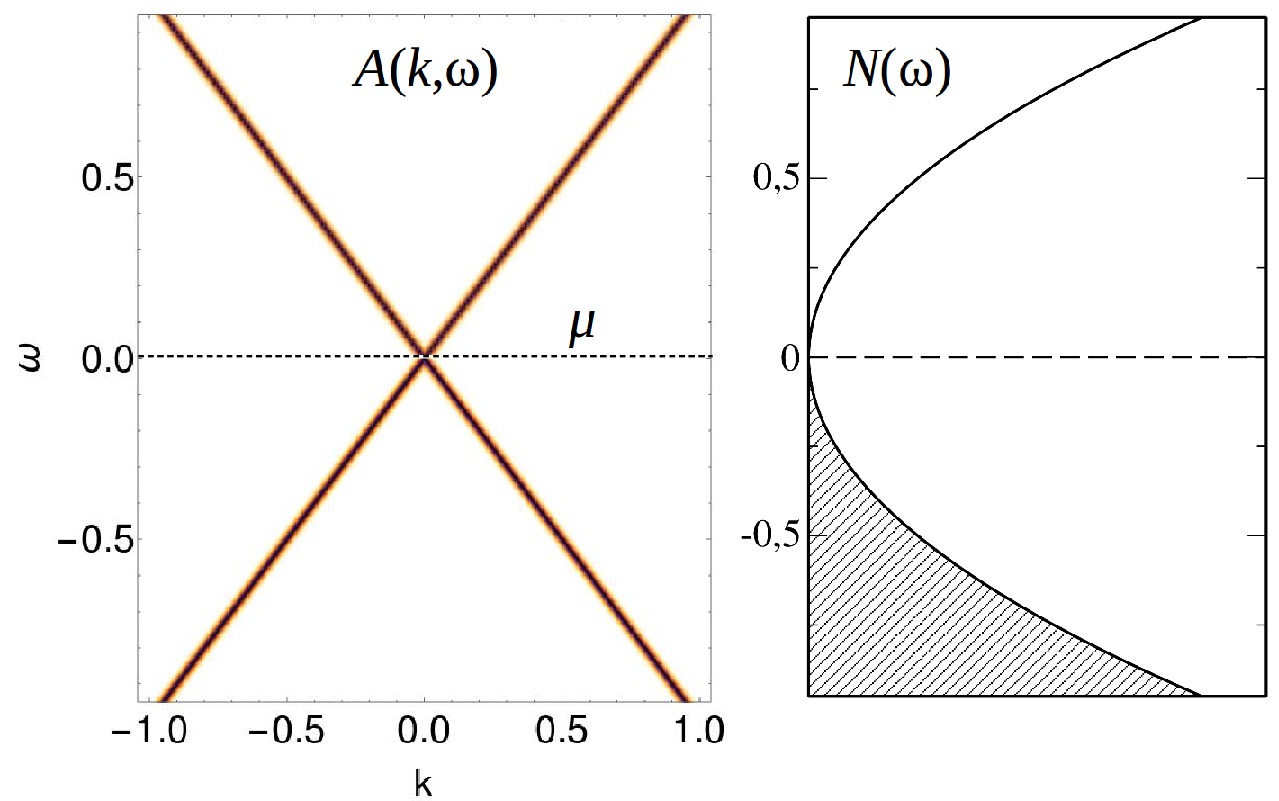}
\caption{Spectral function $A(k,\omega)$ and density of states $N(\omega)$ for a 3D DM in the normal state. Energy and momentum are in dimensionless units.}
\label{fig1a}
\end{figure}

\begin{figure}[ht!]
\centering
\includegraphics[width=0.98\linewidth,clip=true]{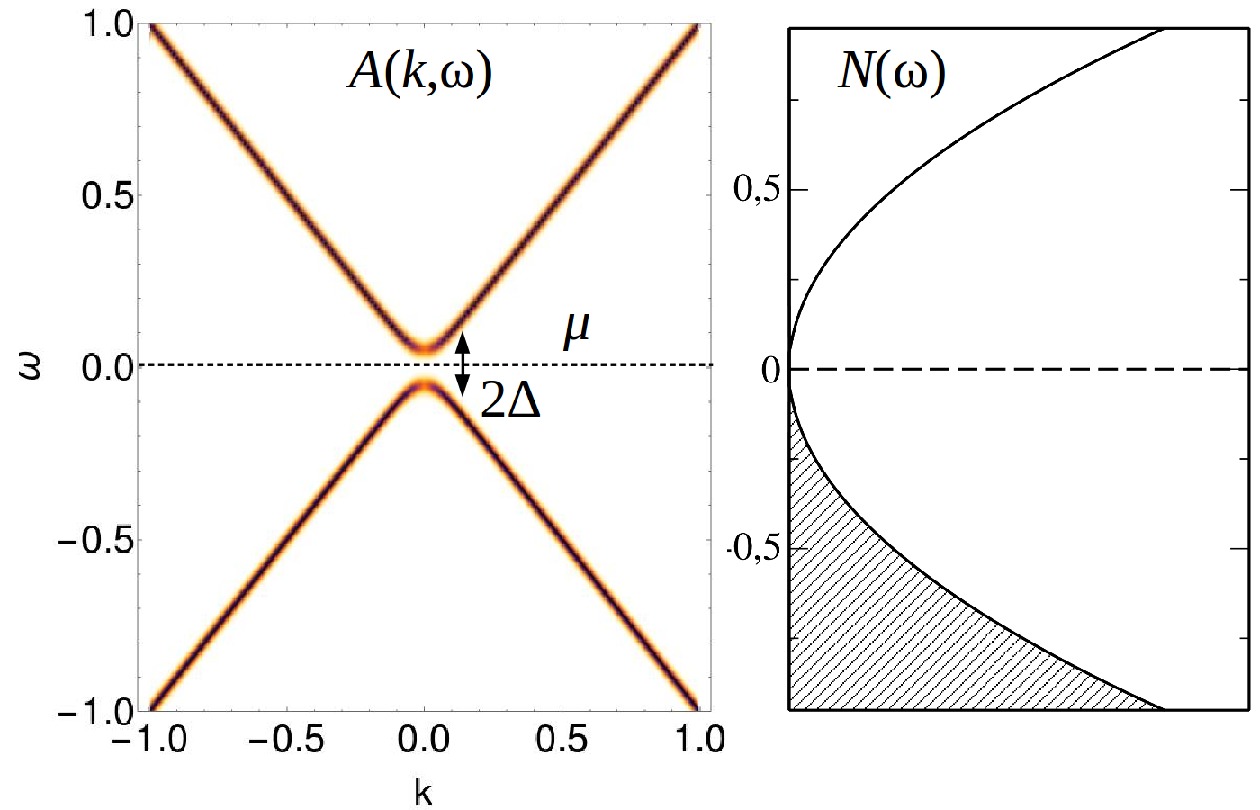}
\caption{Spectral function $A(k,\omega)$ and density of states $N(\omega)$ for an equilibrium  3D DM ($\mu_e=-\mu_h=0$) in the CDW phase for  
 $\alpha=1.7$. 
A gap opens up at the equilibrium chemical potential. 
Energy and momentum are in dimensionless units.}
\label{fig1b}
\end{figure}

The spectral function and the DOS for a single Weyl node in the normal state (no interactions included) and in equilibrium CDW excitonic phase are shown in Fig.~\ref{fig1a} and 
\ref{fig1b}, respectively. The excitonic phase is characterized by a gap which opens up at the equilibrium chemical potential, $\mu=0$. 
In this case, we calculated the gap self-consistently using Eq.~(\ref{eq:CDW_eq_final}) for $\mu_e=-\mu_h=0$~\cite{numerical_note} (see also Appendix~\ref{appendixD}). 
Both EI and CDW phase is accompanied by a gap opening at the Weyl/Dirac node. The excitonic gap is larger in the CDW phase compared to the EI phase.

\begin{figure}[ht!] 
\centering
\includegraphics[width=0.98\linewidth,clip= true]{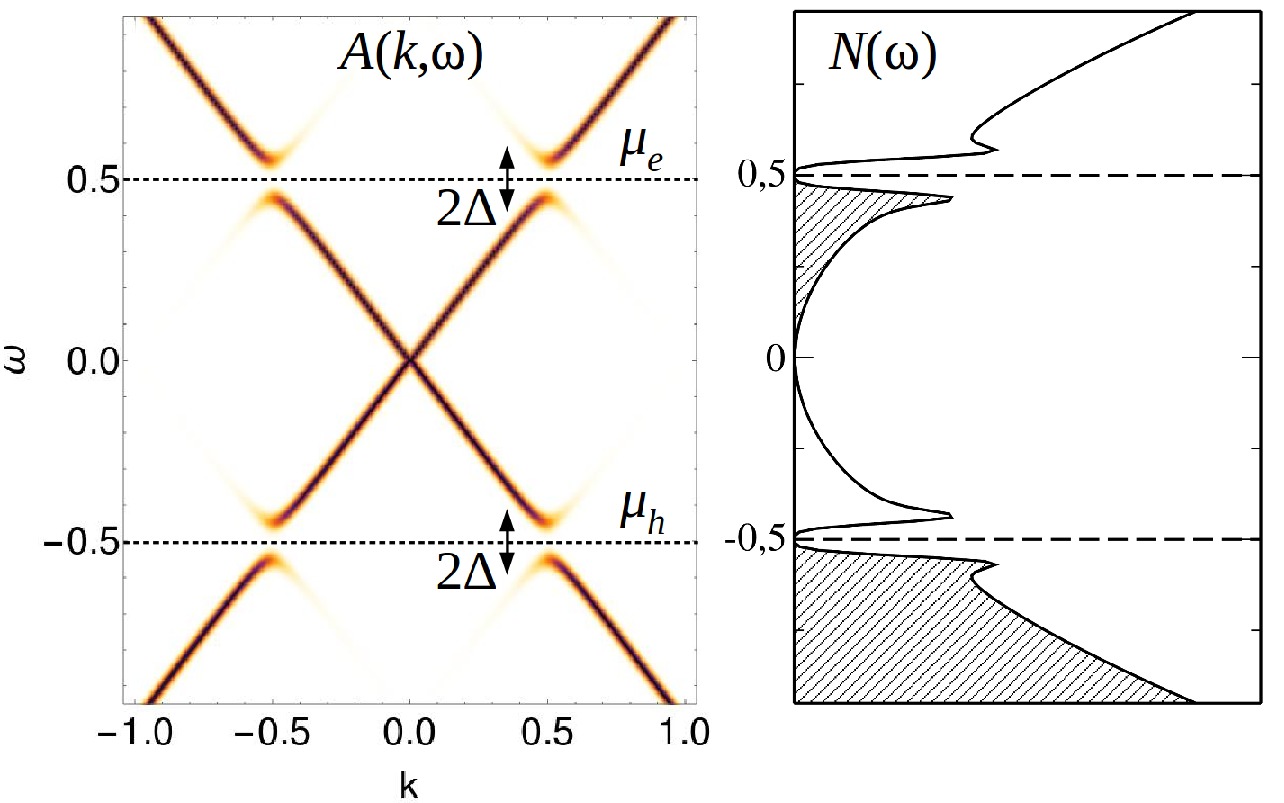}
\caption{Spectral function $A(k,\omega)$ and density of states $N(\omega)$ for a pumped 3D DM ($\mu_e=-\mu_h=0.1$) in the EI phase. 
 For illustrative purposes, in this calculation we use $g=1$ and $\alpha=3$. 
Gaps open up at the quasi-equilibrium chemical potentials $\mu_e$ and $\mu_h$. 
Energy and momentum are in dimensionless units.}
\label{fig1c}
\end{figure}

Figure~\ref{fig1c} shows the spectral function and the DOS in a pumped 3D DM with population inversion. 
In this case, the gap is calculated self-consistently using Eqs.~(\ref{eq:EI_gap_eq_final})-(\ref{eq:CDW_eq_final}) for screened Coulomb potential and  $\mu_e=-\mu_h\ne 0$. 
 At this point, it is important to describe the role of the Dirac cone 
degeneracy $g$. The degeneracy factor is the number of Dirac cones in the system. In the effective model of a 3D DM considered here, $g=2$ for 
a DSM and for a WSM with two nodes, and $g=1$ for a single Weyl node. In real materials the degeneracy can be large, for example $g=24$ in TaAs WSM. 
In this work we take into account metallic screening for situations when the chemical potential is away from the compensation point. This is precisely the case for 
population inversion generated by optical pumping (see Fig.~\ref{fig:0}). The degeneracy factor enters the definition of 
the Thomas-Fermi screening wavevector [see Table~\ref{Table:screening}]. In fact, the screening 
wavevector increases with increasing the chemical potential, the dimensionless coupling constant or the degeneracy factor.  
Therefore screening becomes stronger for larger $g$. 

For optical pumping in a 3D DM, we consider the following cases, (i) all nodes are equally affected by pumping, (ii) pumping is realized selectively for 
 a certain number of nodes. In a WSM with two non-degenerate nodes, the first case corresponds to $g=2$ in the calculation of the screening wavevector while 
the second case to $g=1$. This can be generalized to WSM with the total number of nodes $N$, where $N$ is an even integer, and pumping on $N_0$ nodes, where $N_0\le{N}$. 
At the same time, one needs to keep track of the type of interactions (internodal or intranodal) that are possible in the two cases. 
For $g=2$, both types of interactions are present and therefore both the EI and CDW phases can be realized. For $g=1$, one of the cones has a population inversion, 
with finite Fermi surfaces for electrons and holes, while the other cone is in equilibrium and its corresponding Fermi surface shrinks to a point. The  
strongest pairing is realized for intranodal interactions. Internodal interactions are in principle possible but the resulting excitonic gap vanishes rapidly as a function of 
the mismatch between the equilibrium and non-equilibrium chemical potentials. (The same holds for intranodal interactions with mismatched electron and hole chemical potentials, 
$\mu_e\ne\mu_h$~\cite{triola_prb2017}). In Fig.~\ref{fig1c} we plot the results for the transient EI phase since the gap in the CDW phase is vanishingly small for the present choice of parameters 
($g=1$ and $\mu_e=-\mu_h=0.1$ in dimensionless units). 
In a DSM with the minimal degeneracy $g=2$ only intranodal 
interactions are included, however for $g\ge 4$, both intranodal and internodal 
interactions are possible.

In the pumped 3D DM, the excitonic phase is characterized by gaps that open up at the non-equilibrium electron and hole chemical potentials as illustrated in Fig.~\ref{fig1c}. 
The gaps can be seen in the spectral function, which is indirectly probed in ARPES experiments. In the DOS, the gaps separate occupied and non-occupied states in 
the valence and conduction bands. Such spectroscopic features could be probed by scanning tunneling microscopy.

\subsection{Phase diagrams of the excitonic condensate}\label{Phase_diag}
\begin{figure}[ht!]
\centering
\includegraphics[width=0.98\linewidth,clip=true]{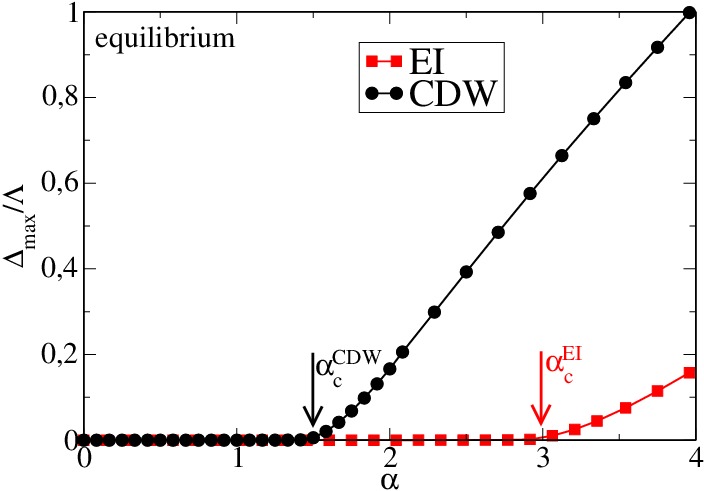}
\caption{Maximum of the gap as a function of the dimensionless coupling $\alpha$ for EI and CDW phase in equilibrium ($\mu_{e}=-\mu_{h}=0$). 
Vertical arrows mark the critical coupling in the two phases. Energy is in dimensionless units.}
\label{fig2a}
\end{figure}

\begin{figure}[ht!]
\centering
\includegraphics[width=0.98\linewidth,clip=true]{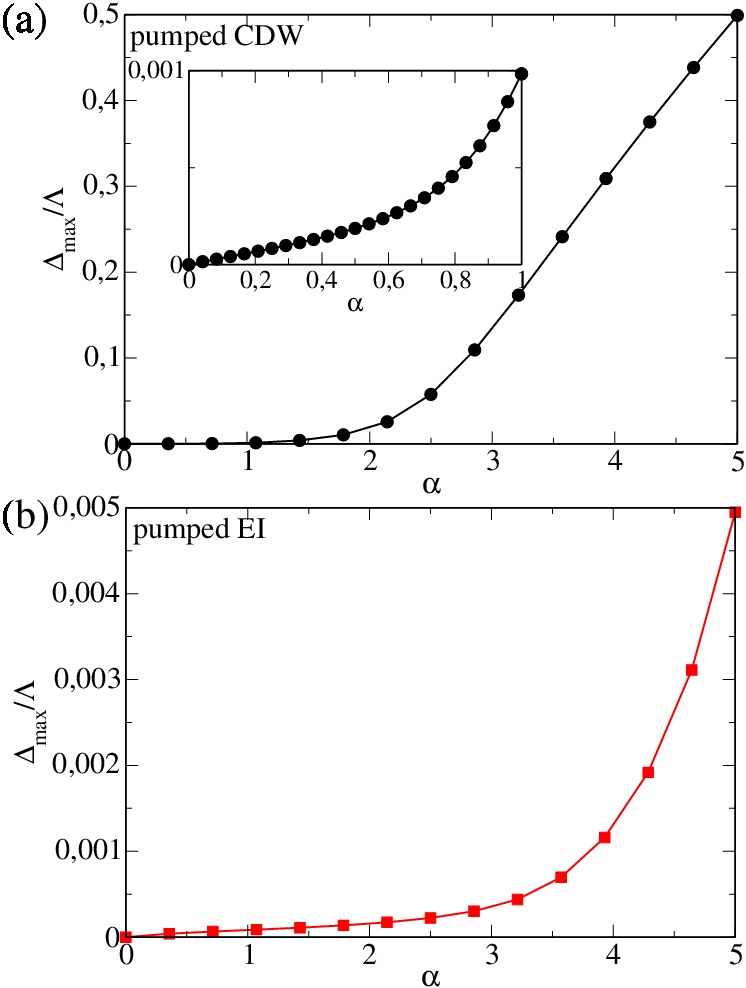}
\caption{Maximum of the gap as a function of the dimensionless coupling $\alpha$ for (a) CDW and (b) EI phase in the pumped regime ($\mu_{e}=-\mu_{h}=0.1$ and $g=2$). 
Inset in panel (a) shows the zoom-in for $0\le\alpha\le1$. Note that in the pumped case the critical coupling vanishes. Energy is in dimensionless units.}
\label{fig2b}
\end{figure}

The dependence of the excitonic gap on the dimensionless coupling constant in equilibrium and in pumped 3D DM is shown in 
Figs.~\ref{fig2a} and \ref{fig2b} respectively. In equilibrium, there is a critical coupling $\alpha_c^{\mathrm{CDW/EI}}$ for CDW(EI) phase such 
that for $\alpha\ge\alpha_c^{\mathrm{CDW(EI)}}$, the gap becomes different from zero. 
The values of the critical coupling that we obtained numerically 
using Eq.~(\ref{eq:EI_gap_eq_final}) and (\ref{eq:CDW_eq_final}) with $\mu_e=-\mu_h=0$~\cite{numerical_note2}, are in agreement with analytical results of Ref.~\cite{Wei_prb2014}.

In the pumped regime with finite chemical potentials for electrons and holes, the critical coupling for excitonic instability vanishes (see Fig.~\ref{fig2b}). 
This numerical result was proven analytically using a model with contact interaction in the case of 2D DM~\cite{triola_prb2017}. Therefore, for any value 
of the coupling $\alpha$, both the CDW and EI phase can be realized in the pumped case. Both the internodal and intranodal interactions contribute 
to the gap opening at the non-equilibrium chemical potentials. 
However, due to the structure of the self-consistent gap equation 
the value of the excitonic gap in the EI phase is always smaller. In our model calculations we are able to consider the two phases separately and to calculate the 
corresponding contributions to the excitonic gap.

\begin{figure}[ht!]
\centering
\includegraphics[width=0.98\linewidth,clip=true]{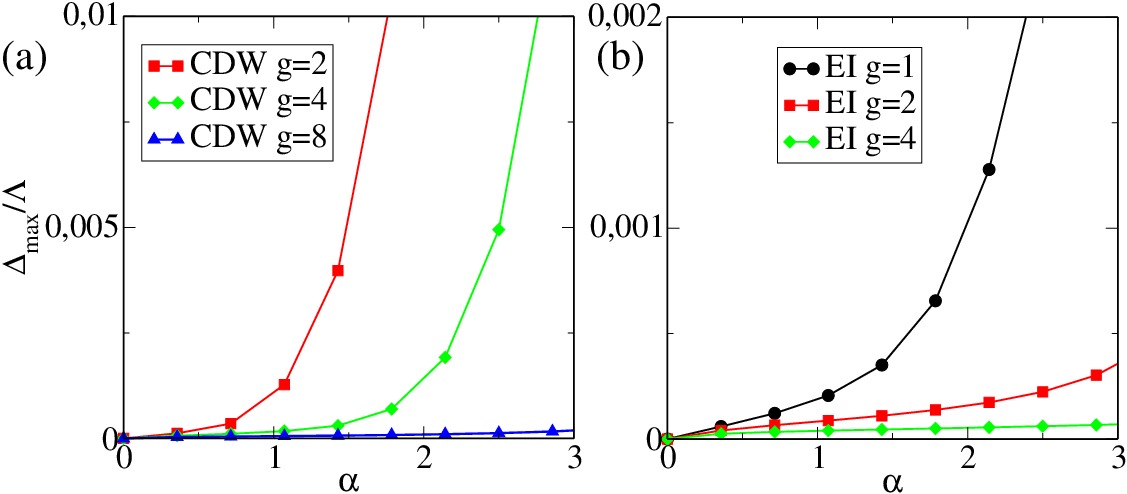}
\caption{Maximum of the gap as a function of the dimensionless coupling $\alpha$ for a pumped 3D DM with $\mu_e=-\mu_h=0.1$ 
in (a) CDW phase with $g=2,4$ and $8$ 
and (b) EI phase with $g=1,2$ and $4$ corresponding to pumping on a single Weyl node, a pair of nodes and two pairs of nodes (see discussion in the main text).   
Energy is in dimensionless units.}
\label{fig3}
\end{figure}

In the equilibrium case with chemical potential exactly at the node, the size of the gap is controlled by the strength of the coupling $\alpha$. 
In the pumped regime, there are two additional factors,  the degeneracy 
 $g$ which affects the screening and the chemical potentials, $\mu_e$ and $\mu_h$, which control both the screening and the DOS at the electron and hole Fermi surfaces. 
 Figure~\ref{fig3} shows the dependence of the gap on $\alpha$ for fixed and equal in magnitude chemical potentials and few different values of $g$. One can see the 
 reduction in the size of the gap with increasing $g$ due to  screening.

\begin{figure}[ht!]
\centering
\includegraphics[width=0.98\linewidth,clip=true]{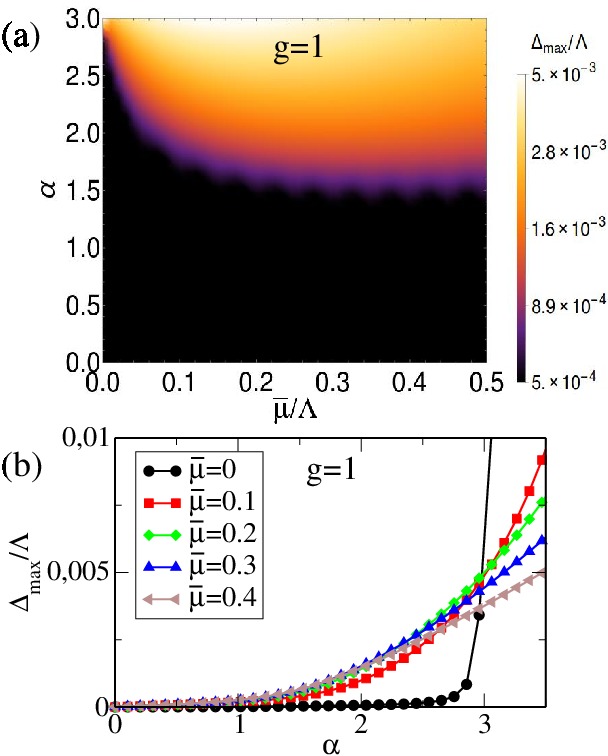}
\caption{(a) Maximum of the gap in the $\alpha$-$\bar{\mu}$ plane for a WSM with $g=1$ (pumping on a single Weyl node) for $0\le\alpha\le{3}$. (b) Maximum of the gap 
as a function of $\alpha$ for different values of $\bar{\mu}$ for $g=1$.  
Energy is in dimensionless units. 
For $g=1$ the main contribution to the gap comes from the EI phase. Critical coupling for the EI phase in 
equilibium ($\bar{\mu}=0$) is $\alpha^{\mathrm{EI}}_{\mathrm{c}}\approx 3.0$ (see also Fig.~\ref{fig2a}). The color shade in panel (a) 
represents the absolute value of the gap and is logarithmically scaled for better contrast.}
\label{fig4a}
\end{figure}

\begin{figure}[ht!]
\centering
\includegraphics[width=0.98\linewidth,clip=true]{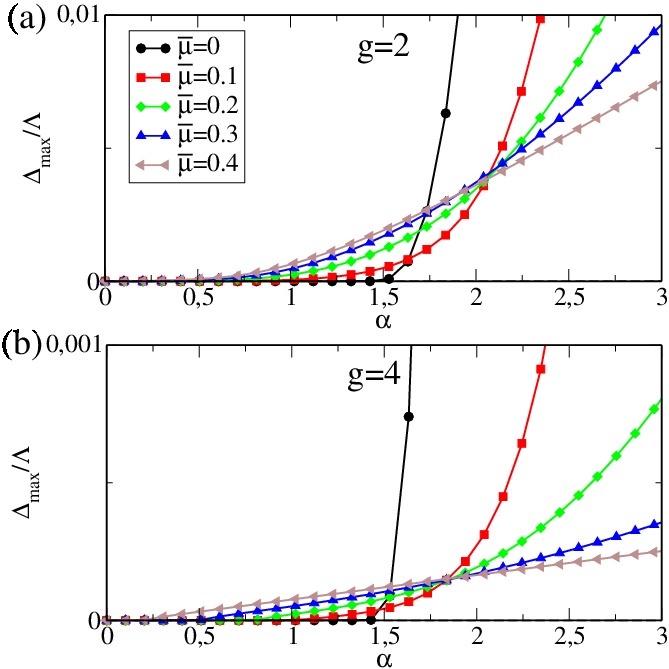}
\caption{Maximum of the gap 
as a function of $\alpha$ for different values of $\bar{\mu}$ for (a) $g=2$ and (b) $g=4$. Other parameters are the same 
as in Fig.~\ref{fig4a}. Energy is in dimensionless units. For $g>1$  the main contribution to the gap comes from the CDW phase. Critical coupling for the CDW phase in 
equilibium ($\bar{\mu}=0$) is $\alpha^{\mathrm{CDW}}_{\mathrm{c}}\approx 1.5$ (see also Fig.~\ref{fig2a}). }
\label{fig4b}
\end{figure}

An important question to answer is whether optical pumping creates more favorable conditions for excitonic instability compared to equilibrium. In principle, in the pumped 
regime excitonic condensation occurs for arbitrary weak coupling strengths due to finite DOS at the electron and hole chemical potentials (see Fig.~\ref{fig2b}). 
However, the size of the gap decreases with increasing the chemical potentials due to screening. As a result of this interplay, pumping is efficient only in a certain segment of the parameter 
space defined by material parameters as illustrated in Figs.~\ref{fig4a} and \ref{fig4b}.  Here we introduce for convenience a single quasi-equilibrium chemical potential $\bar{\mu}$ 
assuming $\mu_e=-\mu_h=\bar{\mu}$. Equal chemical potential for electrons and holes can be realized if 
 the equilibrium chemical potential is at the Dirac point before pumping (see Fig.~\ref{fig:0}). 

The phase diagram in Fig.~\ref{fig4a}(a) shows the maximum of the gap as a function of the chemical potential $\bar{\mu}$ and 
the dimensionless coupling $\alpha$ in the range $0\le\alpha\le{3}$ for a WSM with $g=1$. Clearly, there is a reduction of the critical coupling for $\bar{\mu}>0$. 
By looking at the scans of the phase diagram, $\Delta_\mathrm{max}$ vs $\alpha$ curves for few different $\bar{\mu}$'s in Fig.~\ref{fig4a}(b), one can see that 
 for $\alpha\lesssim{3}$ the gap is zero in equilibrium ($\bar{\mu}=0$) and its size increases monotonically 
 with increasing $\bar{\mu}$. In this segment of the parameter space, 
 optical pumping promotes excitonic instability while the equilibrium system remains gapless. 

For larger $\alpha$ and $\bar{\mu}$, screening becomes stronger. As a result the size of the gap in the pumped regime decreases and becomes smaller than the equilibrium gap [Fig.~\ref{fig4a}(b) for 
$\alpha\gtrsim 3$], provided that $\alpha$ exceeds the critical coupling for excitonic instability in equilibrium. Similar behavior is observed for larger degeneracy factor as shown in Fig.~\ref{fig4b}(a) and (b).
The gap decreases further with increasing $g$. 

\begin{figure}[ht!]
\centering
\includegraphics[width=0.98\linewidth,clip=true]{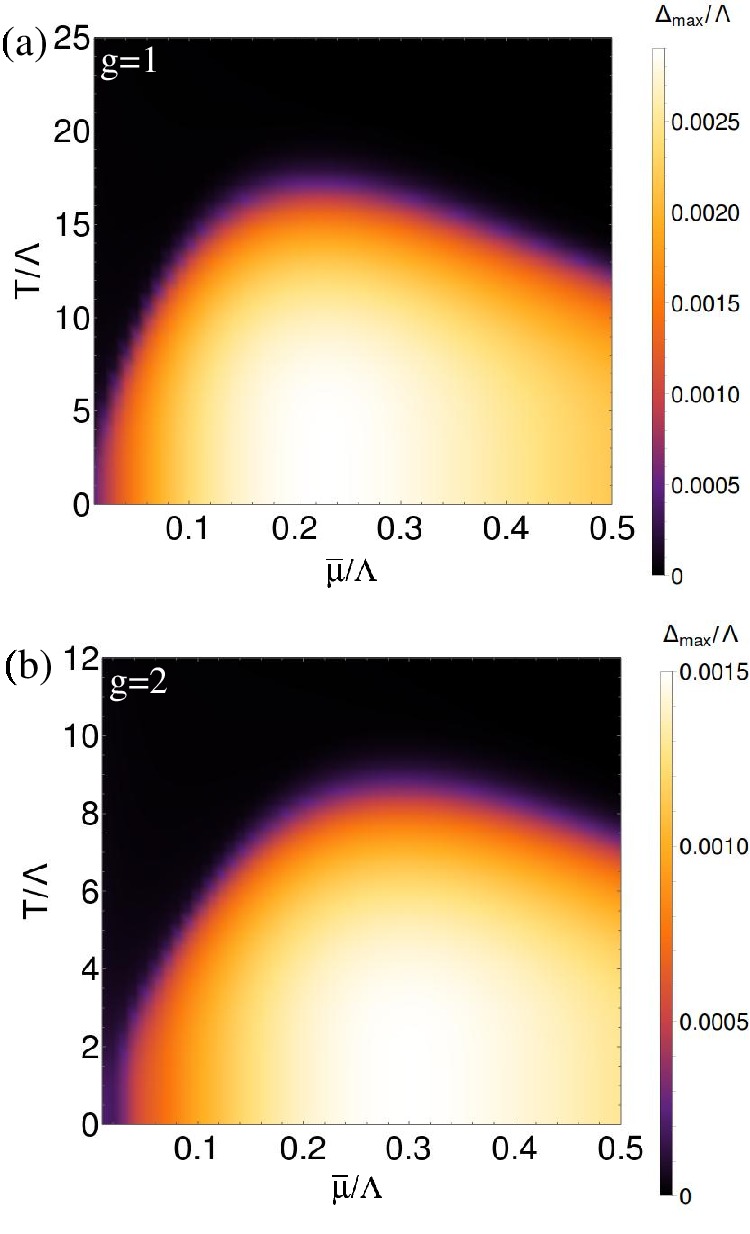}
\caption{Maximum of the gap in $T$-$\bar{\mu}$ plane for a WSM for (a) $g=1$ and $\alpha=2.5$ and (b) $g=2$ and $\alpha=1$. 
Energy and temperature are in dimensionless units. The color shade represents the absolute value of the gap.}
\label{fig5}
\end{figure}

The effect of pumping can also be seen in the behavior of the critical temperature $T_c$ as a function of the chemical potential $\bar{\mu}$. We define $T_c$ as a value of the temperature such 
that for $T\le{T_c}$ the excitonic gap is different from zero. Figure~\ref{fig5} shows the maximum of the gap as a function of $T$ and $\bar{\mu}$ for a WSM with $g=1$  
[Fig.~\ref{fig5}(a)] and $g=2$ [Fig.~\ref{fig5}(b)]. The values of $\alpha$ are chosen to be just below the equilibrium critical coupling for the EI and CDW phases.  This is the 
coupling regime considered in Fig.~\ref{fig4a} and Fig.~\ref{fig4b} in which pumping is most efficient (for the present choice of parameters). 
The line separating the dark 
($\Delta_\mathrm{max}=0$) and bright ($\Delta_\mathrm{max}>0$) regions of the phase diagram defines the dependence of $T_c$ on the chemical potential.  
As shown in Fig.~\ref{fig5}(a), $T_c$ increases with the chemical potential until it reaches a maximum at $\bar{\mu}\approx{0.2}$. Further increasing 
$\bar{\mu}$ leads to reduction of $T_c$ due to screening. Similar behaviour is observed for $g=2$ [see Fig.~\ref{fig5}(b)]; however, $T_c$ reaches a maximum at a slightly 
larger $\bar{\mu}$ due to different values of $g$ and $\alpha$.
%

\subsection{Estimates of critical temperature and excitonic gap}
In order to provide an estimate of the size of the effects proposed in this work, we 
consider some examples of real material realizations of DSM and WSM (see Table~\ref{tab:estimates1}). Anticipating future material discoveries, 
we also consider several examples of 3D DMs with parameters tuned in such a way as to reduce the screening effects and maximize the size of the gap and $T_c$. 
Although many 3D DMs have been proposed in the last few years, here we focus on two examples, Cd$_3$As$_2$  and TaAs, for which extensive ARPES data and detailed density functional 
theory (DFT) calculations 
are available. This allowed us to extract material parameters for numerical calculations. 

There are several important parameters that control the size of the gap and $T_c$, (i) the Dirac cone velocity $v$ and the dielectric constant $\varepsilon$ 
 of the material which determine the value of the 
dimensionless coupling constant $\alpha$, (ii) the energy scale over which the 3D Dirac states exist (the cutoff energy scale $\Lambda$ in our calculations); 
this energy scale limits the range of chemical potentials of the inverted electron/hole populations that can be achieved by pumping, (iii) the Dirac cone degeneracy $g$.  
In the majority of 3D DMs the Dirac dispersion is anisotropic, with the velocity in the $z$-direction ($v_z$) typically different from in-plane velocities ($v_{x,y}$). 
Also the velocities in the upper and lower Dirac cones might differ slightly. For our numerical estimates, we use the average velocity. The Dirac cone degeneracy in real 
materials can be quite large, which is detrimental for excitonic effects 
in pumped 3D DM due to metallic screening. Here we present optimistic estimates assuming selective pumping with small Dirac cone degeneracy that gives the maximum gap and $T_c$ 
($g=2$ for a WSM and $g=4$ for a DSM in the CDW phase). 
For hypothetical 3D DM, we consider a larger range of $g$'s to show the reduction of the gap with increasing the degeneracy. 
\begin{table}[ht!]
\caption{Estimates for $T_c$ and the maximum of the excitonic gap $\Delta_\mathrm{max}$ at $T=0$ for existing and hypothetical 3D DM. 
For  Cd$_3$As$_2$, $v_x\approx v_y\approx 1.3\times 10^{6}$~m/s, $v_z\approx 3.3\times 10^5$~m/s~\cite{liu_natmat2014_dsm},  $\Lambda\approx 1$~eV~\cite{liu_natmat2014_dsm}, 
and  $\varepsilon=36$~\cite{madelung_1998_cd2as2}. For the TaAs family, $v\approx 2.5\times 10^5$~m/s~\cite{lee_prb2015_taas}, $\Lambda\approx 200$~meV~\cite{lee_prb2015_taas,xu2015weyl}, 
and $\varepsilon=10$~\cite{dadsetani_2016_taas}. 
For hypothetical 3D DMs the cutoff energy scale $1$~eV is assumed. Note that for 3D DM with $g=1$, only EI phase is realized; the 
resulting gap and $T_c$ are smaller than in 3D DM with $g=2$ in the CDW phase.}
\begin{tabular}{r||c|c|c|r}
System  & $\alpha$  & $\Lambda$ (eV) & $T_c$ (K) & $\Delta_{\mathrm{max}}$ (meV) \\
\hline
\hline
Cd$_3$As$_2$ DSM &  $0.1$ & $1$ & $0.1$ &  $0.03$ \\
\hline
TaAs WSM &  $1$ & $0.2$ & $2$ &  $0.3$ \\
\hline
\hline
3D DM $g=1$ &  $1-3$ & $1$ & $1-20$ &  $0.3-3$ \\
3D DM $g=2$ &  $1-3$ & $1$ & $10-60$ &  $1-10$ \\
\hline
3D DM $g=4$ &  $1-3$ & $1$ & $1-2$ &  $0.1-0.3$ \\
\hline
\end{tabular}
\label{tab:estimates1}
\end{table}

The results of numerical calculations for $\Delta_\mathrm{max}$ and $T_c$ are summarized in Table~\ref{tab:estimates1}. 
The first two rows correspond to Cd$_3$As$_2$ DSM and TaAs family of WSMs which includes TaAs,TaP,NbAs, and NbP. 
Values of the dimensionless coupling constant for these materials are based on the values of Dirac velocities and dielectric constants found in the literature. 
The energy scale of the 3D Dirac states is extracted from ARPES and DFT (see Table~\ref{tab:estimates1}). 
The rest of the results refer to 3D DMs with theoretical parameters. As one can see from Table~\ref{tab:estimates1}, in TaAs WSM the 
excitonic gap can reach $\approx{0.3}$~meV with critical temperature of few K ($T_c\approx{2}$~K). In Cd$_3$As$_2$, 
we find a gap to be only a small fraction of meV which might be difficult to observe experimentally. 
This is mainly due to relatively large dielectric constant, close to that of a typical 3D TI, which leads to small coupling constant $\alpha$.

Our theory allows us to derive a set of general criteria for material candidates in which transient excitonic instability can be realized and possibly observed 
experimentally. The criteria are (i) large coupling (i.e. small Dirac cone velocity and small dielectric constant), (ii) large $\Lambda$ (the energy extent of 
the Dirac states), and (iii) small Dirac cone degeneracy. This is illustrated by our results for hypothetical 3D DM (see the last three rows in Table~\ref{tab:estimates1})
 where we assumed $\Lambda$ similar to that of graphene and $\alpha\ge{1}$. For degeneracy $g\le{4}$, a gap of tens of meV's and $T_c$ of the order of $100$~K can be achieved.

%

\subsection{Optical conductivity in pumped 3D DM}\label{Opt_cond}
The real part of the frequency-dependent optical conductivity can be calculated using the spectral form of the Kubo formula 
\begin{eqnarray}\label{eq:Kubo}
\mathrm{Re}[\sigma_{\alpha\beta}(\Omega)] &=& \frac{e^2 \pi}{\Omega}\int_{-\infty}^{+\infty}d\omega[n_{F}(\omega)-n_{F}(\omega+\Omega)]\nonumber\\
 &\phantom{=}&\times\sum_{\mathbf{k}}\mathrm{Tr}[\hat{v}_{\alpha}A(\mathbf{k},\omega)v_{\beta}A(\mathbf{k},\omega+\Omega)],
\end{eqnarray}
where $\hat{v}=\frac{\partial H}{\partial \mathbf{k}}$ is the velocity operator and $A(\mathbf{k},\omega)$ is the spectral function. 
In this section we will consider a DM in the normal state, both in equilibrium and with population inversion, and in the exciton phase. 
In the normal state, $A(\mathbf{k},\omega)$ represents the spectral function of the quasiparticles. In the exciton phase, 
$A(\mathbf{k},\omega)$ is the full spectral function, which is a matrix whose 
 diagonal (off-diagonal) elements give the quasiparticle (anomalous) spectral function~\cite{carbotte_prb2017_SC_opt}. 
 In what follows, we only present the final results of the calculations 
 for different cases considered (for details of the calculations see Appendix~\ref{appendixE}.) 

\subsubsection{Equilibrium}\label{Opt_cond_eq}
We start with a brief reminder on the optical properties of non-interacting Dirac fermions in equilibrium (no population inversion). Optical 
conductivity of a WSM have been studied in recent work~\cite{Hosur_prl2012,Timusk_prb2013,Ashby_prb2014,Tabet_prb2016_optical}. 
Here we consider a non-interacting Weyl Hamiltonian for one node with a fixed chirality ($\xi=+1$). For concreteness we assume that the chemical potential $\mu>0$. 
After calculating the spectral function of a WSM (see Appendix~\ref{appendixE}), we find that the optical conductivity is given by
%
%
%
\begin{eqnarray}
\mathrm{Re}[\sigma_{xx}(\Omega)]&=&\frac{e^2}{6\pi v_{F}\Omega}\int_{-\infty}^{+\infty}d\omega[n_{F}(\omega)-n_{F}(\omega+\Omega)]\nonumber\\
&\phantom{=}&\times\int_{0}^{\infty}d\varepsilon\varepsilon^2[A_{+}A_{+}'+A_{-}A_{-}'+2(A_{+}A_{-}'\nonumber\\
&\phantom{=}&\phantom{\times\int_{0}^{\infty}\varepsilon^2[A_{+}A_{+}'+A_{-}A_{-}'}
+A_{-}A_{+}')],\label{eq:opt_WSM}
\end{eqnarray}
%
where $A_{\pm}=\delta(\omega-\varepsilon_{\mathbf{k}}^{\pm})$ and $A_{\pm}'=\delta(\omega+\Omega-\varepsilon_{\mathbf{k}}^{\pm})$.  
The first two terms in the kernel of the energy integral in Eq.~(\ref{eq:opt_WSM}) give the \textit{intraband} contribution to the optical conductivity, while 
the last two terms give the \textit{interband} contribution.

At $T=0$, we obtain the expressions for the intraband and interband conductivity, respectively
\begin{align}
\mathrm{Re}[\sigma_{xx}^\mathrm{intra}(\Omega)] & =\frac{e^2\mu^2}{6\pi v_{F}}\delta(\Omega)\label{eq:opt_WSM_intra}\\
\mathrm{Re}[\sigma_{xx}^\mathrm{inter}(\Omega)] & =\frac{e^2\Omega}{24\pi v_{F}}\Theta(\Omega-2\mu)\label{eq:opt_WSM_inter},
\end{align}
where $\mu$ is the chemical potential of the system and $\Theta(\omega)$ is a step function ($\Theta(\omega)=0$ for $\omega<0$ and 
$\Theta(\omega)=1$ for $\omega>0$).

As one can see from Eq.~(\ref{eq:opt_WSM_inter}), in 3D DM the interband conductivity vanishes linearly with $\Omega$, unlike in graphene 
where the interband piece is constant and equal to $e^2/4\hbar$, the universal optical conductivity of clean graphene. The linear frequency dependence of 
 the optical conductivity of 3D Dirac fermions has been observed experimentally~\cite{Timusk_prb2013}. 

\subsubsection{Pumping: normal state}\label{Opt_cond_pumped}
We will now consider the case of optical pumping in which population inversion is realized. Optical conductivity of 
pumped grapheme with population inversion has been studied in Refs.~\cite{Svintsov_ape2014, Svintsov_optexpress2014}. 
In the case of population inversion, the Fermi-Dirac distributions of conduction band (electrons) and valence band (holes) are given by
\begin{equation}
n^{e/h}_{\mathrm{F}}(\omega)=1/(e^{(\omega-\mu_{e/h})/T}+1),
\end{equation}
where $\mu_{e/h}$ is the corresponding chemical potential. Note that $n^{e}_\mathrm{F}(\omega)$ and $n^{h}_\mathrm{F}(\omega)$ are defined on a segment 
$\omega>0$ and $\omega<0$, respectively. Assuming equal in magnitude chemical potentials for electrons and holes, the 
quasi-equilibrium chemical potential $\bar{\mu}=\mu_e=-\mu_h$. 

\begin{figure}[ht]
\centering
\includegraphics[width=0.98\linewidth,clip=true]{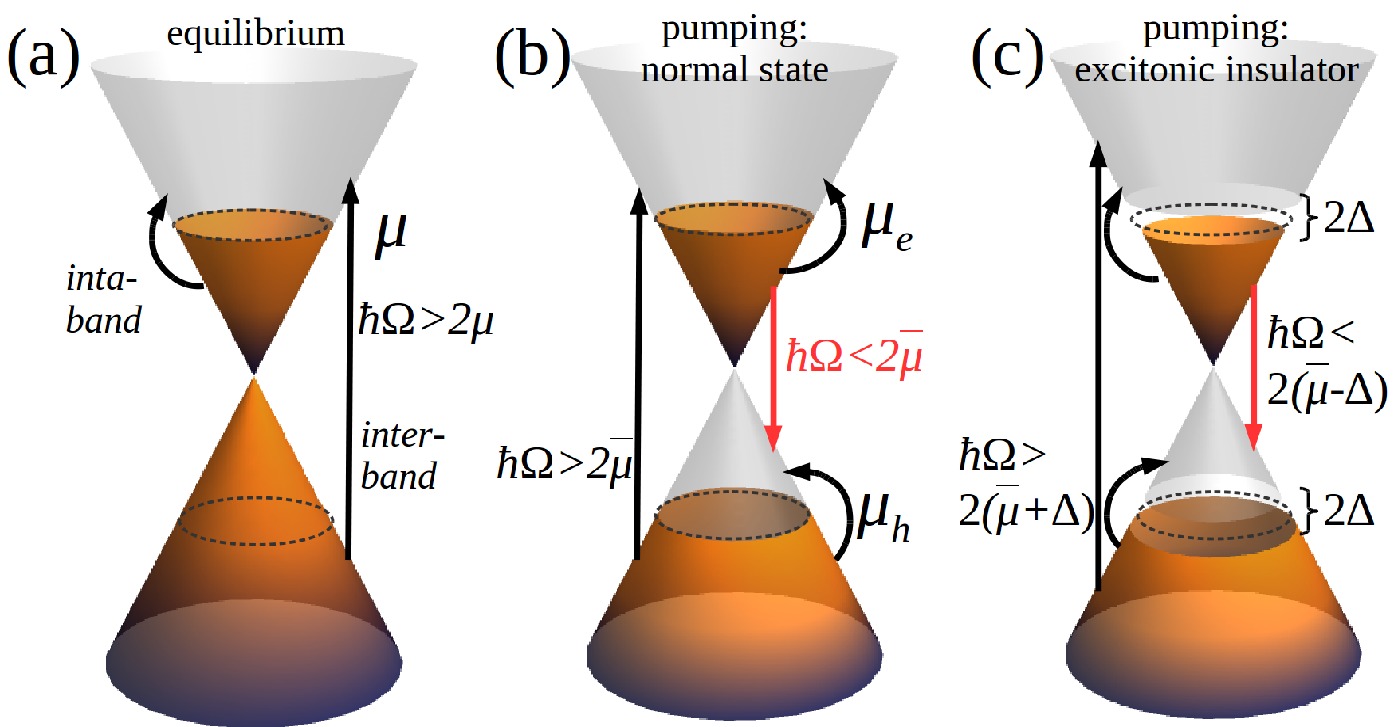}
\caption{Optical transitions in (a) equilibrium DM, (b) pumped DM in the normal state, and (c) pumped DM in 
the excitonic insulator state. $\Delta$ denotes the excitonic gap; $\mu$ and $\bar{\mu}$ are the equilibrium and quasi-equilibrium 
chemical potentials, respectively; $\Omega$ is the frequency of incident light.}
\label{opt_cond_scheme}
\end{figure}

Interband optical conductivity for a pumped WSM with population inversion is given by
\begin{align}
\mathrm{Re}[\sigma_{xx}^\mathrm{inter}(\Omega),T] & =\frac{e^2}{24\pi{v}}[n^{h}_{\mathrm{F}}(-\Omega/2)-n^{e}_{\mathrm{F}}(\Omega/2)].
\end{align}
In the equilibrium case [Eq.~(\ref{eq:opt_WSM_inter})], direct interband transitions are possible only if $\Omega>2\mu$, while 
in the pumped case interband conductivity is positive for $\Omega>2\bar{\mu}$ and negative for $\Omega<2\bar{\mu}$. This is also illustrated in Fig.~\ref{opt_cond_scheme}.
The analytical expressions for optical conductivity for 2D (grapheme) and 3D (single Weyl node) in equilibrium and under population inversion are summarized in 
Table~\ref{tab:opt_cond}. Numerical calculations of the optical conductivity with momentum-dependent gap including quasiparticle and anomalous contributions 
will be considered elsewhere.

\begin{table*}[ht!]
\caption{Interband optical conductivity of 2D and 3D DM  (graphene and WSM, respectively) in equilibrium 
and under population inversion. In equilibrium 
the system is described by a single Fermi-Dirac distribution 
 $n^{0}_\mathrm{F}(\omega)$, while in the pumped case there are two separate Fermi-Dirac distribution 
 for valence and conduction band, $n^{h}_\mathrm{F}(\omega)$ and $n^{e}_\mathrm{F}(\omega)$, 
respectively. Here $\sgn(\omega)$ is a sign function, $\sgn(\omega)=1$ ($-1$) for $\omega>0$ ($\omega<0$). The proper 
units of optical conductivity are restored by including $\hbar$ in the expressions for $\mathrm{Re}[\sigma(\Omega)]$.}
\begin{tabular}{r||c|c|r}
System  & Equilibrium  & Pumping \\

\hline
\hline
2D DM & 
\begin{tabular}{rc} $T\ne0:\quad$ & $\frac{e^2}{4\hbar}[n^{0}_{\mathrm{F}}(-\Omega/2)-n^{0}_{\mathrm{F}}(\Omega/2)]$ \\  
                   $T=0:\quad$ & $\frac{e^2}{4\hbar}\Theta(\Omega-2\mu)$
\end{tabular} 
& 
\begin{tabular}{rc} $T\ne0:\quad$ & $\frac{e^2}{4\hbar}[n^{h}_{\mathrm{F}}(-\Omega/2)-n^{e}_{\mathrm{F}}(\Omega/2)]$  \\  
                   $T=0:\quad$ & $\frac{e^2}{4\hbar}\,\sgn(\Omega-2\mu)$ 
\end{tabular}\\
\hline
3D DM & 
\begin{tabular}{rc} $T\ne0:\quad$ & $\frac{e^2}{24\pi\hbar^2{v}_{F}}\Omega[n^{0}_{\mathrm{F}}(-\Omega/2)-n^{0}_{\mathrm{F}}(\Omega/2)]$ \\  
                   $T=0:\quad$ & $\frac{e^2}{24\pi\hbar^2{v}_{F}}\Omega\Theta(\Omega-2\bar{\mu})$
\end{tabular}     &  
\begin{tabular}{rc} $T\ne0:\quad$ & $\frac{e^2}{24\pi\hbar^2{v}_{F}}\Omega[n^{h}_{\mathrm{F}}(-\Omega/2)-n^{e}_{\mathrm{F}}(\Omega/2)]$  \\  
                   $T=0:\quad$ & $\frac{e^2}{24\pi\hbar^2{v}_{F}}\Omega\,\sgn(\Omega-2\bar{\mu})$
                   \end{tabular}
\end{tabular}
\label{tab:opt_cond}
\end{table*}

\subsubsection{Pumping: excitonic phase}\label{Opt_cond_exc}
Optical conductivity of a superconductor or an excitonic insulator can be computed using the general Kubo formula [Eq.~(\ref{eq:Kubo})] 
using the full spectral function of the system which includes quasiparticle and anomalous contributions. 
 In the case of a superconductor, the full Green's function, and therefore the spectral function, can be conveniently 
written as a matrix in the Nambu basis (see Appendix~\ref{appE_order_bcs}). In the case of an excitonic insulator, 
it can be written as a matrix in the electron/hole basis (see Appendix~\ref{appE_order_ex})
\begin{eqnarray}
 \mathcal{A}(\mathbf{k},\omega)=\left(
 \begin{array}{lc}
  A_e(\mathbf{k},\omega)  & B(\mathbf{k},\omega) \\
  B^{\dagger}(\mathbf{k},\omega)  & A_h(\mathbf{k},\omega) \\
 \end{array}
 \right),
\end{eqnarray}
where 
\begin{eqnarray}
A_e(\mathbf{k},\omega) & = & u_{e\mathbf{k}}^2\delta(\omega-\omega_{+})+v_{e\mathbf{k}}^2\delta(\omega-\omega_{-}),\label{eq:EI_Ac}\\
A_h(\mathbf{k},\omega) & = & u_{h\mathbf{k}}^2\delta(\omega-\omega_{+})+v_{h\mathbf{k}}^2\delta(\omega-\omega_{-}),\label{eq:EI_Av}\\
B(\mathbf{k},\omega) & = & -\frac{\Delta_{\mathbf{k}}}{\omega_{+}-\omega_{-}}[\delta(\omega-\omega_{+})-\delta(\omega-\omega_{-}].\label{eq:EI_B}
\end{eqnarray}
The spectral weights are given by
\begin{eqnarray}
\begin{array}{lc}
u^2_{e\mathbf{k}}  = \frac{\omega_{+}-\varepsilon^{-}_{\mathbf{k}}}{\omega_{+}-\omega_{-}}, & v^2_{e\mathbf{k}} = \frac{\omega_{-}-\varepsilon^{-}_{\mathbf{k}}}{\omega_{-}-\omega_{+}},\\
u^2_{h\mathbf{k}}  = \frac{\omega_{+}-\varepsilon^{+}_{\mathbf{k}}}{\omega_{+}-\omega_{-}}, & v^2_{h\mathbf{k}} = \frac{\omega_{-}-\varepsilon^{+}_{\mathbf{k}}}{\omega_{-}-\omega_{+}}.
\end{array}
\end{eqnarray}

The optical conductivity of a pumped 3D DM in the excitonic phase reads
%
%
%
\begin{widetext}
\begin{eqnarray}\label{eq:opt_cond_pump}
\mathrm{Re}[\sigma_{xx}^\mathrm{EX}(\Omega)]
&=&\frac{e^2\pi}{\Omega}\int_{-\infty}^{+\infty}d\omega[n_{F}(\omega)-n_{F}(\omega+\Omega)]
\sum_{\mathbf{k}}[(\omega^{+}_{k_x})^2 A_e(\mathbf{k},\omega)A_e(\mathbf{k},\omega+\Omega) +(\omega^{-}_{k_x})^2 A_h(\mathbf{k},\omega)A_h(\mathbf{k},\omega+\Omega)\nonumber\\
&\phantom{=}&\phantom{\frac{e^2\pi}{\Omega}\int_{-\infty}^{+\infty}d\omega[n_{F}(\omega)-n_{F}(\omega+\Omega)]
\sum_{\mathbf{k}}[}+\omega^{+}_{k_x}\omega^{-}_{k_x}(B(\mathbf{k},\omega)B^{\dagger}(\mathbf{k},\omega+\Omega)+B^{\dagger}(\mathbf{k},\omega)B(\mathbf{k},\omega+\Omega))],
\end{eqnarray}
\end{widetext}
where we introduced
\begin{eqnarray}
\hat{\omega}_{k_{\alpha}}=\Bigg(
\begin{array}{lc}
\omega^{+}_{k_{\alpha}}  & 0 \\
0 &  \omega^{-}_{k_{\alpha}}
\end{array}
\Bigg),
\end{eqnarray}
and $\omega^{\pm}_{k_{\alpha}}=\partial\omega_{\pm}/\partial{k_{\alpha}}$. Here $\varepsilon_{\mathbf{k}}^{\pm}$ are the electron/hole dispersions 
[Eq.~(\ref{bands_pumped_CB})-(\ref{bands_pumped_VB})] and $\omega_{\pm}$ 
are the excitonic bands [Eq.~(\ref{exc_bands})]. The order parameter $\Delta_{\mathbf{k}}$ can be calculated self-consistently using Eqs.~(\ref{eq:EI_gap_eq_final})-(\ref{eq:CDW_eq_final}). 
In general, the optical conductivity should be calculated numerically. However, assuming a homogeneous gap ($\Delta_{\mathrm{k}}=\Delta_0$), one can obtain useful analytical results. 
Focusing on the quasiparticle contribution, we find that the intraband conductivity in the valence and conduction bands is proportional to $\Theta(\Omega-2\Delta_0)$, while the 
interaband conductivity is proportional to $\Theta(\Omega-2(\bar{\mu}+\Delta_0))$. This is illustrated schematically in Fig.~\ref{opt_cond_scheme}(c).

\section{Conclusions}\label{concl}
In conclusion, we have proposed to search for transient excitonic states in optically-pumped 3D Dirac materials. 
Such states are characterized by gaps in the quasiparticle spectrum which open up at the non-equilibrium chemical potentials of photoexcited electrons and holes. 
 In the case of pumped Weyl semimetals, two possible excitonic phases exist at arbitrary weak interaction strength, the excitonic insulator 
 phase and the charge density wave phase which originate from intranodal and internodal interactions respectively. Both types of interactions contribute to the 
 gap opening away from the node. We have calculated the phase diagrams of the transient 
 excitonic condensate that result from the interplay between the enhancement of the density of states at the electron and hole Fermi surfaces and the screening of the Coulomb interaction. 
 We have found that there exist a region of the parameter space defined by the dimensionless coupling constant and the Dirac cone degeneracy in which 
 optical pumping is more favorable for excitonic condensation compared to equilibrium. 
 
 Numerical calculations have shown that in some of the existing 
 materials, excitonic gaps of the order of $1$~meV and critical temperatures of few K can be achieved. Following the general criteria 
 for enhancement of the gap and $T_c$,  we have demonstrated that much lager effect can be found in materials with tuned parameters, e.g. 
 large coupling constant and small Dirac cone degeneracy. Considering the fast rate of materials prediction and discovery, such parameters could 
 be realized in novel 3D DMs.  Finally, we have identified the electronic and optical signatures of the transient excitonic condensate that can be 
probed experimentally. Given the growing interest in non-equilibrium Dirac matter and the increasing capabilities of time-resolved spectroscopic pump-probe techniques, 
we anticipate that transient many-body states in Dirac materials will become an important topic. 
 
\section*{Acknowledgement}
This work was supported by ERC-DM-32031, KAW, CQM. Work at LANL was supported by USDOE BES E3B7. We acknowledge support 
from Dr. Max R{\"{o}}ssler, the Walter Haefner Foundation and the ETH Zurich Foundation, and the Villum Center for Dirac Materials. We 
 are grateful to P. Hofmann, C. Triola, D. Abergel, K. Dunnett and S. Banerjee for useful discussions
\appendix
\section{Eigenstates of the non-interacting Weyl Hamiltonian}\label{appendixA}
Consider a WSM described by a pair of nodes, one with chirality $\xi=+1$ located at 
 $\mathbf{q}=\mathbf{k}-\mathbf{K}$ and the other one with chirality $\xi=-1$ located at $\mathbf{q}=\mathbf{k}+\mathbf{K}$ [$K\ne{0}$ 
 and $K_0=0$ in Eq.~(\ref{eq:HW}) of the main text]. We will refer to the two nodes 
 as the right (R) and the left (L) one. Then the Hamiltonian of the R/L node is given by
\begin{eqnarray}
H_{\pm}(\mathbf{k}\mp\mathbf{K})&\equiv & H_{R/L}(\mathbf{q}),\\
H_{R/L}(\mathbf{q})&=&\pm\hbar v\mathbf{\sigma}\cdot\mathbf{q}.\label{eq:a1_HW_LR}
\end{eqnarray}
The eigenstates of the Hamiltonian in Eq.~(\ref{eq:a1_HW_LR}) are obtained by solving the corresponding Schr\"{o}dinger equation
\begin{eqnarray}
H_{R/L}\chi_{\mathbf{q}}^{R/L}&=E\chi_{\mathbf{q}}^{R/L}\label{eq:a1_Weyl_SE},
\end{eqnarray}
where $\chi_{\mathbf{q}}^{R/L}$ are the eigenvectors. 
There are two eigenvalues, $\varepsilon_{\mathbf{q}}^{+}=+\hbar{v}|\mathbf{q}|$ (conduction band) and $\varepsilon_{\mathbf{q}}^{-}=-\hbar{v}|\mathbf{q}|$ (valence band). 
The corresponding normalized eigenvectors are given by two component spinors $\chi_{\mathbf{q},\pm}^{L/R}$
%
\begin{eqnarray}
\chi_{\mathbf{q},+}^{R}=
\left(
\begin{array}{l}
 \cos{\frac{\theta}{2}}e^{-i\phi} \\ 
 \sin{\frac{\theta}{2}}
\end{array}
\right),
\chi_{\mathbf{q},-}^{R}=
\left(
\begin{array}{l}
 -\sin{\frac{\theta}{2}}e^{-i\phi}\label{eq:a1_Weyl_eigenv1}\\ 
 \cos{\frac{\theta}{2}}
\end{array}
\right)\\
\chi_{\mathbf{q},+}^{L}=
\left(
\begin{array}{l}
 -\sin{\frac{\theta}{2}}e^{-i\phi} \\ 
 \cos{\frac{\theta}{2}}
\end{array}
\right),
\chi_{\mathbf{q},-}^{L}=
\left(
\begin{array}{l}
 \cos{\frac{\theta}{2}}e^{-i\phi} \\ 
 \sin{\frac{\theta}{2}}
\end{array}
\right).\label{eq:a1_Weyl_eigenv2}
\end{eqnarray}
%

From the eigenvectors $\chi_{\mathbf{q},\pm}^{L/R}$ we can construct a unitary transformation $\tilde{H}_{R/L}(\mathbf{q})={U^{R/L}}^{\dagger}H_{R/L}(\mathbf{q}))U^{R/L}$ 
that diagonalizes the Hamiltonian $H_{R/L}(\mathbf{q})$, where the matrix $U^{R/L}$ given by
\begin{equation}\label{eq:Weyl_unitary_transform}
 U^{R/L}=\left(\begin{array}{lr}
  \chi_{\mathbf{q},+}^{R/L,\uparrow} & \chi_{\mathbf{q},-}^{R/L,\uparrow} \\
  \chi_{\mathbf{q},+}^{R/L,\downarrow} & \chi_{\mathbf{q},-}^{R/L,\downarrow}
 \end{array}
 \right).
\end{equation}

The Hamiltonian in the diagonal basis reads
\begin{equation}\label{eq:Weyl_diag}
 \tilde{H}_{R/L}=\sum_{\mathbf{q}}{c_{\mathbf{q}}^{R/L}}^{\dagger}\tilde{H}_{R/L}(\mathbf{q})c_{\mathbf{q}}^{R/L}, 
\end{equation}
where
\begin{eqnarray}
 c_{\mathbf{q}}^{R/L}&=&\left(
\begin{array}{l}
 c_{\mathbf{q},+}^{R/L} \\
 c_{\mathbf{q},-}^{R/L}
\end{array}
 \right),\\
 \tilde{H}^{R/L}(\mathbf{q})&=&\left(\begin{array}{lr}
  \varepsilon_{\mathbf{q}}^{+} & 0 \\
  0 & \varepsilon_{\mathbf{q}}^{-}
 \end{array}
 \right).
\end{eqnarray}
Here ${c_{\mathbf{q},\pm}^{R/L}}^{\dagger}(c_{\mathbf{q},\pm}^{R/L})$ are the fermionic creation(annihilation) operators corresponding to the two bands. 

Furthermore, we can re-write the spinors $\Phi_{\mathbf{q}}^{R/L}$ in the new basis
\begin{equation}
 \Phi_{\mathbf{q}}^{R/L}=U^{R/L}c_{\mathbf{q}}^{R/L},
\end{equation}
or more explicitly
\begin{equation}
 \Phi_{\mathbf{q}}^{R/L}=\chi_{\mathbf{q},+}^{R/L} c_{\mathbf{q},+}^{R/L}+\chi_{\mathbf{q},-}^{R/L} c_{\mathbf{q},-}^{R/L}.
\end{equation}
The last expression can be re-written in the following form
\begin{eqnarray}
 \Phi_{\mathbf{q}}^{R/L}&=&\Phi_{\mathbf{q},+}^{R/L}+\Phi_{\mathbf{q},-}^{R/L}=\sum_{n=\pm}\Phi_{\mathbf{q},n}^{R/L},\\
 \Phi_{\mathbf{q},n}^{R/L}&\equiv&\chi_{\mathbf{q},n}^{R/L} c_{\mathbf{q},n}^{R/L}.
\end{eqnarray}
The corresponding expressions for the spin-components of $\Phi_{\mathbf{q}}^{R/L}$ are given by
\begin{eqnarray}
 \Phi_{\mathbf{q},\sigma}^{R/L}&=&\sum_{n=\pm}\Phi_{\mathbf{q},n,\sigma}^{R/L},\label{Phi_sigma}\\
 \Phi_{\mathbf{q},n,\sigma}^{R/L}&\equiv&\chi_{\mathbf{q},n}^{R/L,\sigma} c_{\mathbf{q},n}^{R/L}\label{Phi_n_sigma},
\end{eqnarray}
where $\sigma=\uparrow,\downarrow$ denotes the spin of the state and $\chi_{\mathbf{q},n}^{R/L,\sigma}$ are the spin-components of the 
eigenvectors in Eqs.~(\ref{eq:a1_Weyl_eigenv1}-\ref{eq:a1_Weyl_eigenv2}).

\section{Derivation of the interacting Weyl Hamiltonian}\label{appendixB}
The general spin-independent particle-particle interaction for a system of two Weyl nodes can be written as
\begin{equation}
 V=\sum_{\sigma,\sigma^{\prime}}\sum_{\mathbf{k},\mathbf{k}',\mathbf{q}}\sum_{\alpha_i=L/R}{\Phi_{\mathbf{k}'+\mathbf{q},\sigma'}^{\alpha_1{\dagger}}}
 \Phi_{\mathbf{k}',\sigma'}^{\alpha_2} {\Phi_{\mathbf{k}-\mathbf{q},\sigma}^{\alpha_3{\dagger}}}
 \Phi_{\mathbf{k},\sigma}^{\alpha_4}.
\end{equation}

Taking into account the conservation of energy, after some manipulations and relabeling, we get
\begin{widetext}
\begin{eqnarray}
V&=&\sum_{\sigma,\sigma^{\prime}}\sum_{\mathbf{k},\mathbf{k}'}\left\lbrace\right.
 V(0)({\Phi_{\mathbf{k}',\sigma'}^{L}}^{\dagger}\Phi_{\mathbf{k}',\sigma'}^{L}+{\Phi_{\mathbf{k}',\sigma'}^{R}}^{\dagger}\Phi_{\mathbf{k}',\sigma'}^{R})
 ({\Phi_{\mathbf{k},\sigma}^{L}}^{\dagger}\Phi_{\mathbf{k},\sigma}^{L}+{\Phi_{\mathbf{k},\sigma}^{R}}^{\dagger}\Phi_{\mathbf{k},\sigma}^{R})
+V(\mathbf{k}'-\mathbf{k})({\Phi_{\mathbf{k},\sigma'}^{L}}^{\dagger}\Phi_{\mathbf{k}',\sigma'}^{L}{\Phi_{\mathbf{k}',\sigma}^{L}}^{\dagger}\Phi_{\mathbf{k},\sigma}^{L}\nonumber\\
&\phantom{=}&\phantom{\sum_{\sigma,\sigma^{\prime}}\sum_{\mathbf{k},\mathbf{k}'}\left\lbrace\right.}
+{\Phi_{\mathbf{k},\sigma'}^{R}}^{\dagger}\Phi_{\mathbf{k}',\sigma'}^{R}{\Phi_{\mathbf{k}',\sigma}^{R}}^{\dagger}\Phi_{\mathbf{k},\sigma}^{R}
+{\Phi_{\mathbf{k},\sigma'}^{R}}^{\dagger}\Phi_{\mathbf{k}',\sigma'}^{L}{\Phi_{\mathbf{k}',\sigma}^{L}}^{\dagger}\Phi_{\mathbf{k},\sigma}^{R}+
 {\Phi_{\mathbf{k},\sigma'}^{L}}^{\dagger}\Phi_{\mathbf{k}',\sigma'}^{R}{\Phi_{\mathbf{k}',\sigma}^{R}}^{\dagger}\Phi_{\mathbf{k},\sigma}^{L})\nonumber\\
 &\phantom{=}&\phantom{\sum_{\sigma,\sigma^{\prime}}\sum_{\mathbf{k},\mathbf{k}'}\left\lbrace\right.}
 +V(2\mathbf{K})({\Phi_{\mathbf{k}'+2\mathbf{K},\sigma'}^{R}}^{\dagger}\Phi_{\mathbf{k}',\sigma'}^{L}{\Phi_{\mathbf{k}-2\mathbf{K},\sigma}^{L}}^{\dagger}\Phi_{\mathbf{k},\sigma}^{R})
 +V(-2\mathbf{K})({\Phi_{\mathbf{k}'-2\mathbf{K},\sigma'}^{L}}^{\dagger}\Phi_{\mathbf{k}',\sigma'}^{R}{\Phi_{\mathbf{k}+2\mathbf{K},\sigma}^{R}}^{\dagger}\Phi_{\mathbf{k},\sigma}^{L})\nonumber\\
 &\phantom{=}&\phantom{\sum_{\sigma,\sigma^{\prime}}\sum_{\mathbf{k},\mathbf{k}'}\left\lbrace\right.}
 +V(\mathbf{k}-\mathbf{k}'+2\mathbf{K})({\Phi_{\mathbf{k}+2\mathbf{K},\sigma'}^{R}}^{\dagger}\Phi_{\mathbf{k}',\sigma'}^{R}{\Phi_{\mathbf{k}'-2\mathbf{K},\sigma}^{L}}^{\dagger}\Phi_{\mathbf{k},\sigma}^{L})
  +V(\mathbf{k}-\mathbf{k}'-2\mathbf{K})({\Phi_{\mathbf{k}-2\mathbf{K},\sigma'}^{L}}^{\dagger}\Phi_{\mathbf{k}',\sigma'}^{L}{\Phi_{\mathbf{k}'+2\mathbf{K},\sigma}^{R}}^{\dagger}\Phi_{\mathbf{k},\sigma}^{R})
 \left.\right\rbrace.\nonumber\\
\end{eqnarray}
\end{widetext}

Next, noticing that the term proportional to $V(0)$ vanishes and taking into account that $V(\mathbf{k})=V(-\mathbf{k})$, the interaction term becomes
%
\begin{eqnarray}
V&=&\sum_{\sigma,\sigma^{\prime}}\sum_{\mathbf{k},\mathbf{k}'}\left\lbrace\right.
 V(\mathbf{k}'-\mathbf{k})({\Phi_{\mathbf{k},\sigma'}^{L}}^{\dagger}\Phi_{\mathbf{k}',\sigma'}^{L}{\Phi_{\mathbf{k}',\sigma}^{L}}^{\dagger}\Phi_{\mathbf{k},\sigma}^{L}\nonumber\\
&+&{\Phi_{\mathbf{k},\sigma'}^{R}}^{\dagger}\Phi_{\mathbf{k}',\sigma'}^{R}{\Phi_{\mathbf{k}',\sigma}^{R}}^{\dagger}\Phi_{\mathbf{k},\sigma}^{R}+
2{\Phi_{\mathbf{k},\sigma'}^{L}}^{\dagger}\Phi_{\mathbf{k}',\sigma'}^{R}{\Phi_{\mathbf{k}',\sigma}^{R}}^{\dagger}\Phi_{\mathbf{k},\sigma}^{L})\nonumber\\
%
&+&2V(2\mathbf{K})({\Phi_{\mathbf{k}'+2\mathbf{K},\sigma'}^{R}}^{\dagger}\Phi_{\mathbf{k}',\sigma'}^{L}{\Phi_{\mathbf{k}-2\mathbf{K},\sigma}^{L}}^{\dagger}\Phi_{\mathbf{k},\sigma}^{R})\nonumber\\
&+&2V(\mathbf{k}-\mathbf{k}'-2\mathbf{K})({\Phi_{\mathbf{k}-2\mathbf{K},\sigma'}^{L}}^{\dagger}\Phi_{\mathbf{k}',\sigma'}^{L}
{\Phi_{\mathbf{k}'+2\mathbf{K},\sigma}^{R}}^{\dagger}\Phi_{\mathbf{k},\sigma}^{R})
 \left.\right\rbrace.\nonumber\\
\end{eqnarray}

Adopting the notations, $\mathbf{k}\rightarrow\mathbf{q}$, where $q=\mathbf{k\pm\mathbf{K}}$ for R/L node, and using  
$\Phi_{\mathbf{q},\sigma}^{R/L}=\sum_{n=\pm}\Phi_{\mathbf{q,n,\sigma}}^{R/L}$ [Eq.~(\ref{Phi_sigma})], we obtain
\begin{eqnarray}
V&=&-\sum_{\substack{\sigma\sigma^{\prime},\mathbf{q}\mathbf{q}' \\ n=\pm }}\left\lbrace\right.
V(\mathbf{q}'-\mathbf{q})({\Phi_{\mathbf{q},n_1,\sigma'}^{L}}^{\dagger}\Phi_{\mathbf{q},n_2,\sigma}^{L}{\Phi_{\mathbf{q}',n_3,\sigma}^{L}}^{\dagger}\Phi_{\mathbf{q}',n_4,\sigma'}^{L}\nonumber\\
&+&{\Phi_{\mathbf{q},n_1,\sigma'}^{R}}^{\dagger}\Phi_{\mathbf{q},n_2,\sigma}^{R}{\Phi_{\mathbf{q}',n_3,\sigma}^{R}}^{\dagger}\Phi_{\mathbf{q}',n_4,\sigma'}^{R})\nonumber\\
&+&2V(\mathbf{q}-\mathbf{q}'-2\mathbf{K}){\Phi_{\mathbf{q},n_1,\sigma'}^{L}}^{\dagger}\Phi_{\mathbf{q},n_2,\sigma}^{L}{\Phi_{\mathbf{q}',n_3,\sigma}^{R}}^{\dagger}\Phi_{\mathbf{q}',n_4,\sigma'}^{R}\nonumber\\
&-&2V(2\mathbf{K}){\Phi_{\mathbf{q}',n_1,\sigma'}^{R}}^{\dagger}\Phi_{\mathbf{q}',n_2,\sigma'}^{L}{\Phi_{\mathbf{q},n_3,\sigma}^{L}}^{\dagger}\Phi_{\mathbf{q},n_4,\sigma}^{R}\nonumber\\
&+&2V(\mathbf{q}-\mathbf{q}'){\Phi_{\mathbf{q},n_1,\sigma'}^{L}}^{\dagger}\Phi_{\mathbf{q},n_2,\sigma}^{R}{\Phi_{\mathbf{q}',n_3,\sigma}^{R}}^{\dagger}\Phi_{\mathbf{q}',n_4,\sigma'}^{R}
\left.\right\rbrace.
\label{V_Coulomb}
\end{eqnarray}
The first two terms in Eq.~(\ref{V_Coulomb}) correspond to intranodal scattering while the last two terms correspond to internodal 
scattering.

The spin-components of the wavefunctions can be written in the diagonal basis,  
$\sum_{\sigma}\Phi_{\mathbf{q,n,\sigma}}^{R/L}=\chi_{\mathbf{q},n}^{R/L}c_{\mathbf{q},n}^{R/L}$ [Eq.~(\ref{Phi_n_sigma})]. 
After calculating the inner products of the form $\chi_{\mathbf{q},n}^{\alpha}\chi_{\mathbf{q'},-n}^{\beta}$, we arrive at the following expression 
 %
\begin{eqnarray}
V&=&-\sum_{\mathbf{q},\mathbf{q}'}\left\lbrace\right.
 V(\mathbf{q}-\mathbf{q}')\left[\right. 
 A_1(\mathbf{q},\mathbf{q}')(c_{\mathbf{q},+}^{L\dagger}c_{\mathbf{q},-}^L c_{\mathbf{q}',-}^{L\dagger}c_{\mathbf{q}',+}^L\nonumber\\
 &\phantom{=}&\phantom{-\sum_{\mathbf{q},\mathbf{q}'}}
 +c_{\mathbf{q},+}^{R\dagger}c_{\mathbf{q},-}^R c_{\mathbf{q}',-}^{R\dagger}c_{\mathbf{q}',+}^R)+A_2(\mathbf{q},\mathbf{q}')\nonumber\\ 
&\phantom{=}&\phantom{-\sum_{\mathbf{q},\mathbf{q}'}}
(c_{\mathbf{q},-}^{L\dagger}c_{\mathbf{q},+}^L c_{\mathbf{q}',+}^{L\dagger}c_{\mathbf{q}',-}^L
+c_{\mathbf{q},-}^{R\dagger}c_{\mathbf{q},+}^R c_{\mathbf{q}',+}^{R\dagger}c_{\mathbf{q}',-}^R
)\left.\right]\nonumber\\
&+&V(\mathbf{q}-\mathbf{q}'-2\mathbf{K})\left[\right.
B_1(\mathbf{q},\mathbf{q}')c_{\mathbf{q},+}^{L\dagger}c_{\mathbf{q},-}^L c_{\mathbf{q}',-}^{R\dagger}c_{\mathbf{q}',+}^R\nonumber\\
&\phantom{+}&\phantom{V(\mathbf{q}-\mathbf{q}'-2\mathbf{K})}
+B_2(\mathbf{q},\mathbf{q}')c_{\mathbf{q},-}^{L\dagger}c_{\mathbf{q},+}^L c_{\mathbf{q}',+}^{R\dagger}c_{\mathbf{q}',-}^R
\left.\right]\nonumber\\
&-&2V(2\mathbf{K})(c_{\mathbf{q}',+}^{R\dagger}c_{\mathbf{q}',-}^L c_{\mathbf{q},-}^{L\dagger}c_{\mathbf{q},+}^R+
c_{\mathbf{q}',-}^{R\dagger}c_{\mathbf{q}',+}^L c_{\mathbf{q},+}^{L\dagger}c_{\mathbf{q},-}^R)\nonumber\\
&+&2V(\mathbf{q}-\mathbf{q}')C_1(\mathbf{q},\mathbf{q}')(c_{\mathbf{q},+}^{L\dagger}c_{\mathbf{q},-}^R c_{\mathbf{q}',-}^{R\dagger}c_{\mathbf{q}',+}^L\nonumber\\
&\phantom{+}&\phantom{2V(\mathbf{q}-\mathbf{q}')C_1(\mathbf{q},\mathbf{q}')}
+c_{\mathbf{q},-}^{L\dagger}c_{\mathbf{q},+}^R c_{\mathbf{q}',+}^{R\dagger}c_{\mathbf{q}',-}^L)
 \left.\right\rbrace\label{V_Coulomb_final},
\end{eqnarray}
%
where
\begin{eqnarray}
 A_1(\mathbf{q},\mathbf{q}')&=&\frac{\hat{e}_{q}\cdot\hat{e}_{q'}^{*}+\hat{e}_q^{*}\cdot \hat{e}_{q'}}{4}+i\frac{(\hat{q}+\hat{q}')\cdot(\hat{e}_q^2\times\hat{e}_{q'}^{2})}{2}\nonumber\\
 \\
 A_2(\mathbf{q},\mathbf{q}')&=&\frac{\hat{e}_{q}\cdot\hat{e}_{q'}^{*}+\hat{e}_q^{*}\cdot \hat{e}_{q'}}{4}-i\frac{(\hat{q}+\hat{q}')\cdot(\hat{e}_q^2\times\hat{e}_{q'}^{2})}{2}\nonumber\\
 \\
 B_1(\mathbf{q},\mathbf{q}')&=&\frac{\hat{e}_{q}\cdot\hat{e}_{q'}+\hat{e}_q^{*}\cdot \hat{e}_{q'}^{*}}{4}-i\frac{(\hat{q}-\hat{q}')\cdot(\hat{e}_q^2\times\hat{e}_{q'}^{2})}{2}\nonumber\\
 \\
 B_2(\mathbf{q},\mathbf{q}')&=&\frac{\hat{e}_{q}\cdot\hat{e}_{q'}+\hat{e}_q^{*}\cdot \hat{e}_{q'}^{*}}{4}+i\frac{(\hat{q}-\hat{q}')\cdot(\hat{e}_q^2\times\hat{e}_{q'}^{2})}{2}\nonumber\\
 \\
 C_1(\mathbf{q},\mathbf{q}')&=&\frac{\hat{q}\cdot\hat{q}'+1}{2}.
\end{eqnarray}
As before, the last two terms in Eq.~(\ref{V_Coulomb_final}) are the internodal scattering terms. 
Here we work in the spherical coordinate system $\left\lbrace \hat{q},\hat{e}_q^1,\hat{e}_q^2\right\rbrace\equiv\left\lbrace \hat{r},\hat{\theta},\hat{\phi}\right\rbrace$, 
where $\theta$ and $\phi$ are the azimuthal and polar angles, respectively;   
$\hat{e}_q=\hat{e}_q^1+i\hat{e}_{q}^2$. 

Finally, after further manipulations we obtain a concise expression for the interaction potential
\begin{eqnarray}
V&=&-\sum_{\substack{\mathbf{q},\mathbf{q}' \\ n=\pm}}\left\lbrace\right.
 V(\mathbf{q}-\mathbf{q}')A(\mathbf{q},\mathbf{q}')\sum_{\alpha=R,L}c_{\mathbf{q},n}^{\alpha\dagger}c_{\mathbf{q},-n}^{\alpha}c_{\mathbf{q}',-n}^{\alpha\dagger}c_{\mathbf{q}',n}^{\alpha}\nonumber\\
 &\phantom{=}&+V(\mathbf{q}-\mathbf{q}'-2\mathbf{K})B(\mathbf{q},\mathbf{q}')
 c_{\mathbf{q},n}^{L\dagger}c_{\mathbf{q},-n}^{L}c_{\mathbf{q}',-n}^{R\dagger}c_{\mathbf{q}',n}^{R}\nonumber\\
 &\phantom{=}&-\left[2V(2\mathbf{K})-V(\mathbf{q}-\mathbf{q}')C(\mathbf{q},\mathbf{q}')\right]
  c_{\mathbf{q},n}^{L\dagger}c_{\mathbf{q},-n}^{R}c_{\mathbf{q}',-n}^{R\dagger}c_{\mathbf{q}',n}^{L},
 \nonumber\\\label{eq:a2_V_Coulomb_final2}
\end{eqnarray}
where the first term refers to intranodal interactions and the last two terms to internodal interactions. 
Coefficients $A,B,C$ are defined as
\begin{eqnarray}
 A(\mathbf{q},\mathbf{q}')&=&\frac{\hat{e}_{q}\cdot\hat{e}_{q'}^{*}+\hat{e}_q^{*}\cdot \hat{e}_{q'}}{4}\\
 B(\mathbf{q},\mathbf{q}')&=&\frac{\hat{e}_{q}\cdot\hat{e}_{q'}+\hat{e}_q^{*}\cdot \hat{e}_{q'}^{*}}{2}\\
 C(\mathbf{q},\mathbf{q}')&=&\hat{q}\cdot\hat{q}'+1.
\end{eqnarray}

Using the following identities
%
\begin{eqnarray}
 \frac{\hat{e}_{q}\cdot\hat{e}_{q'}^{*}+\hat{e}_q^{*}\cdot \hat{e}_{q'}}{4}&=&\frac{\sin\theta\sin\theta'}{2}\nonumber\\
  &+&\frac{1+\cos\theta\cos\theta'}{2}\cos(\phi-\phi')\nonumber\\
\\
 \frac{\hat{e}_{q}\cdot\hat{e}_{q'}+\hat{e}_q^{*}\cdot \hat{e}_{q'}^{*}}{4}&=&\frac{\sin\theta\sin\theta'}{2}\nonumber\\
&-&\frac{1-\cos\theta\cos\theta'}{2}\cos(\phi-\phi')\nonumber\\
\\ 
 -\frac{(\hat{q}+\hat{q}')\cdot(\hat{e}_q^2\times\hat{e}_{q'}^{2})}{2}&=&\frac{\cos\theta+\cos\theta'}{2}\sin(\phi-\phi')\nonumber\\
\\ 
 -\frac{(\hat{q}-\hat{q}')\cdot(\hat{e}_q^2\times\hat{e}_{q'}^{2})}{2}&=&\frac{\cos\theta-\cos\theta'}{2}\sin(\phi-\phi')\nonumber\\
\\ 
 \frac{\hat{q}\cdot\hat{q}'+1}{2}&=&\frac{1+\cos\theta\cos\theta'}{2}\nonumber\\
 &+&\frac{\sin\theta\sin\theta'}{2}\cos(\phi-\phi'),
\end{eqnarray}
%
we obtain Eq.~(\ref{eq:V_Coulomb}) of the main text.

\section{Thomas-Fermi screening in 2D and 3D}\label{appendixC}

The screening wavevector in the Thomas-Fermi approximation in 2D and 3D is given by, respectively 
\begin{eqnarray}
\kappa_{\mathrm{2D}}&=&\frac{2\pi e^2}{\varepsilon}\frac{\partial n}{\partial \mu},\label{eq:a3_scr_wv_2D}\\
\kappa_{\mathrm{3D}}&=&\sqrt{\frac{4\pi e^2}{\varepsilon}\frac{\partial n}{\partial \mu}}.\label{eq:a3_scr_wv_3D}
\end{eqnarray}
In order to calculate $\kappa_{\mathrm{2D/3D}}$, one needs to know the electron density as a function 
of the chemical potential $\mu$. We will consider systems with linear ($E=\hbar{v}k$) and parabolic ($E=\hbar^2{k^2}/2m$) dispersion 
in 2D and 3D. 

In the case of parabolic dispersion, the Fermi wavevector is given by $k_{\mathrm{F}}=\sqrt{2m\mu}/\hbar$. 
For 2D electron gas (2DEG), the total number of quantum states is given by 
$N=g\frac{A}{(2\pi)^2}\pi{k_{\mathrm{F}}^2}$, where $g$ is the degeneracy factor (spin) and $A$ is the area. 
Then the density is $n=N/A=\frac{{g}k_{\mathrm{F}}^2}{4\pi}=\frac{gm}{2\pi\hbar^{2}}\mu$. Analogously, for 3D electron gas (3DEG), 
$N=g\frac{V}{(2\pi)^3}\frac{4\pi}{3}{{k_{\mathrm{F}}}^3}$, where $V$ is the volume, and  
 $n=N/V=\frac{{g}k_{\mathrm{F}}^3}{6\pi^2}=\frac{gm}{6\pi^2\hbar^{3}}(\sqrt{2{m}\mu})^3$. Substituting expressions for 
 electron density into Eqs.~(\ref{eq:a3_scr_wv_2D}) and (\ref{eq:a3_scr_wv_3D}), we obtain the screening wavevector 
 for 2DEG and 3DEG, respectively
\begin{eqnarray}
\kappa_\mathrm{2DEG}&=&\frac{g e^2 m}{\varepsilon\hbar^2},\label{eq:a3_scr_wv_2DEG}\\
\kappa_\mathrm{3DEG}&=&\sqrt{\frac{2{g}e^2{m}}{\varepsilon\pi\hbar^2}k_{\mathrm{F}}}.\label{eq:a3_scr_wv_3DEG}
\end{eqnarray}

In the case of the Dirac dispersion, the Fermi wavevector is given by $k_{\mathrm{F}}={\mu}/\hbar{v}$. 
For 2D DM, the electron density is $n=N/A=\frac{{g}k_{\mathrm{F}}^2}{4\pi}=\frac{g}{4\pi{v}^2\hbar^{2}}\mu^2$. For 3D DM, 
 $n=N/V=\frac{{g}k_{\mathrm{F}}^3}{6\pi^2}=\frac{g}{6\pi^2{v}^3\hbar^{3}}\mu^3$. Substituting expressions for 
 electron density into Eqs.~(\ref{eq:a3_scr_wv_2D}) and (\ref{eq:a3_scr_wv_3D}), we obtain the screening wavevector 
 for 2D DM and 3D DM, respectively
\begin{eqnarray}
\kappa_\mathrm{2DDM}&=&\frac{g e^2}{\varepsilon{v}\hbar}k_\mathrm{F}\equiv{g}\alpha k_\mathrm{F},\label{eq:a3_scr_wv_2DDM}\\
\kappa_\mathrm{3DDM}&=&\sqrt{\frac{2 g e^2}{\pi\varepsilon{v}\hbar}}k_\mathrm{F}\equiv\sqrt{\frac{2{g}\alpha}{\pi}} k_\mathrm{F}.\label{eq:a3_scr_wv_3DDM}
\end{eqnarray}
The results are summarized in Table ~\ref{Table:A3:screening}.

\begin{table}[ht!]
\caption{Electron density $n$ as a function of the chemical potential, density of states $D(E)$ as a function of energy 
and the Thomas-Fermi screening wavevector $\kappa_\mathrm{TF}$ 
for a Dirac spectrum and electron gas in 2D and 3D. 
}
\begin{tabular}{r||c|c|c}
System  & $n$ &  $D(E)$  & $\kappa_\mathrm{TF}$ \\
\hline
\hline
2D DM & $\frac{g}{4\pi{v}^2\hbar^{2}}\mu^2$  & $\frac{g}{2\pi(\hbar v)^2}E$ & $g\alpha k_\mathrm{F}$\\
\hline
3D DM &  $\frac{g}{6\pi^2{v}^3\hbar^{3}}\mu^3$ &  $\frac{g}{2\pi^2(\hbar v)^3}E^2$ & $\sqrt{\frac{2 g\alpha}{\pi}}k_\mathrm{F}$ \\
\hline
\hline
2DEG  & $\frac{gm}{2\pi\hbar^{2}}\mu^2$ & $\frac{gm}{2\pi\hbar^2}$ & $\frac{{g}e^2{m}}{\varepsilon\hbar^2}$ \\
\hline
3DEG  & $\frac{gm}{6\pi^2\hbar^{3}}(\sqrt{2{m}\mu})^3$ & $\frac{g\sqrt{2m^3}}{{2\pi^2\hbar^3}}E^{1/2}$ & $\sqrt{\frac{2{g}{e^2}{m}}{\varepsilon\pi\hbar^2}} k_\mathrm{F}^{1/2}$ \\
\end{tabular}
\label{Table:A3:screening}
\end{table}

\section{Derivation of the gap equation for pumped 3D DM}\label{appendixD}
In order to derive the gap equation in the general form [Eq.~(\ref{gap_pumped}) in the main text], we consider a simple two-band Hamiltonian of 
a pumped 3D DM in a long-lived quasi-equilibrium state
\begin{eqnarray}
H&=&\sum_{\substack{\mathbf{q} \\ n=\pm}}\varepsilon_{\mathbf{q}}^{n}c_{\mathbf{q},n}^{\dagger}c_{\mathbf{q},n}+
\sum_{\textbf{q},\textbf{k},\textbf{k}'} \tilde{V}({\textbf{q}}) c^{\dagger}_{\textbf{k}+\textbf{q},+}c^{\dagger}_{\textbf{k}^\prime-
\textbf{q},-}c_{\textbf{k}',-}c_{\textbf{k},+},\nonumber\\
\label{eq:H}
\end{eqnarray}
where the first term is the kinetic energy and the last term is the interaction between electrons and holes; 
$\varepsilon^{+}_{\mathbf{k}}=\hbar{v}|\mathbf{k}|-\mu_e$ ($\varepsilon^{+}_{\mathbf{k}}=-\hbar{v}|\mathbf{k}|-\mu_h$) is electron (hole) dispersion 
measured from the 
electron (hole) chemical potential, $\mu_\mathrm{\mathrm{e}}$ ($\mu_\mathrm{\mathrm{h}}$); 
$c_{\mathbf{k},\pm}^{\dagger}$ ($c_{\mathbf{k},\pm}$)  creates (annihilates) 
an electron in band $\varepsilon_{\mathbf{k}}^{\pm}$ with momentum $\textbf{k}$. 
Here we consider a single Dirac node in 3D momentum space and we do not specify the form of the interaction potential $\tilde{V}(\mathbf{q})$. As shown in Sections~\ref{Exc_equil_EI} 
and \ref{Exc_equil_CDW}, the Hamiltonian in Eq.~(\ref{eq:H}) can be easily adjusted to the case of two nodes and intra- or internodal interactions  

We define the electron(hole) Green's functions and the anomalous Green's function as follows (see also Section~\ref{appE_order_ex})
\begin{eqnarray}
G_{e(h)}(\mathbf{k},\tau-\tau') & = & -<T_{\tau} c_{\mathbf{k},\pm}(\tau)c^{\dagger}_{\mathbf{k},\pm}(\tau')>,\\
F(\mathbf{k},\tau-\tau') & = & -<T_{\tau} c_{\mathbf{k},+}(\tau){c_{\mathbf{k},-}}^{\dagger}(\tau')>,
\end{eqnarray}
where $T_{\tau}$ is the imaginary time-ordering operator. The mean-field order parameter, or excitonic gap, is defined as 
\begin{eqnarray}
 \Delta_{\textbf{k}}&=&\sum_{\textbf{k}'}\tilde{V}({\textbf{k}-\textbf{k}'})\langle c_{\textbf{k}',+}c^\dagger_{\textbf{k}',-} \rangle\nonumber\\
 &=&T\sum_{\textbf{k}',i\omega_n}V_{\textbf{k}-\textbf{k}'}F(\textbf{k}';i\omega_n),
\end{eqnarray}
where $\omega_n=(2n+1)\pi/\beta$ is a fermionic Matsubara frequency, $\beta=1/k_{\mathrm{B}}T$ and $T$ is the temperature of electrons and holes.  
We can then proceed with the standard derivation using, for example, the Gor'kov approach~\cite{excitons}, in which time-dependent equations of motions for $G_{e(h)}$ and $F$ are derived. 
From equations of motion and the definition of $\Delta_{\mathbf{k}}$, we get
\begin{equation}
 \Delta(\mathbf{k})=\sum_{\mathbf{k}'}\tilde{V}(\mathbf{k},\mathbf{k}')\frac{\Delta(\mathbf{k}')}{\omega_{+}({\mathbf{k}'})-\omega_{-}({\mathbf{k}'})}
 [n_{\mathrm{F}}(\omega_{+})-n_{\mathrm{F}}(\omega_{-})],
\end{equation}
where 
\begin{equation}
 \omega_{\pm}(\mathbf{k})=\frac{\varepsilon^{+}_{\mathbf{k}}+\varepsilon^{-}_{\mathbf{k}}}{2}\pm\frac{1}{2}
 \sqrt{(\varepsilon^{+}_{\mathbf{k}}-\varepsilon^{-}_{\mathbf{k}})^2+4|\Delta(\mathbf{k})|^2}
\end{equation}
are the renormalized bands and $n_{\mathrm{F}}(\omega)=1/(e^{\beta\omega}+1)$ is the Fermi-Dirac distribution.  

In equilibrium, taking the limit $\mu_e=-\mu_h=0$ and using the properties of the Fermi-Dirac distribution, the gap equation becomes 
\begin{equation}
 \Delta_\mathbf{k}=\sum_{\mathbf{k}'}\tilde{V}(\mathbf{k},\mathbf{k}')\frac{\Delta_\mathbf{k}'}{2E_{\mathbf{k}'}}\tanh{\frac{\beta{E_{\mathbf{k}'}}}{2}},
\end{equation}
where $E_{\mathbf{k}'}=\sqrt{(\hbar{v}|\mathbf{k'}|)^2+|\Delta_\mathbf{k}'|^2}$.

Finally, we comment on the numerical solution of the gap equation with screened Coulomb interaction. As shown in the main text, 
for the case of internodal interactions the gap equation (in dimensionless units) can be written as 
\begin{align}
  \Delta^{\alpha}(\mathbf{q})&=\frac{4\pi{\alpha}}{(2\pi)^3}\int\frac{1}{(\mathbf{q}-\mathbf{q}')^2+\kappa^2}\frac{\Delta^{\alpha}(\mathbf{q}')\cos(\phi-\phi')}
  {\omega_{+}({\mathbf{q}'})-\omega_{-}({\mathbf{q}'})}\nonumber\\
  &\phantom{=}\phantom{\frac{4\pi{\alpha}}{(2\pi)^3}\int\frac{1}{(\mathbf{q}-\mathbf{q}')^2+\kappa^2}}\times[n_{\mathrm{F}}(\omega_{+})-n_{\mathrm{F}}(\omega_{-})]dV,
\end{align}
%
where $dV=|\mathbf{q}'|^2 d{q}'\sin\theta'{d}\theta'{d}\phi'$ and $\kappa$ is the screening wavevector. 
For an $s$-wave order parameter, the Coulomb potential $V({\mathbf{q},\mathbf{q}'})\propto\frac{1}{(\mathbf{q}-\mathbf{q}')^2+\kappa^2}$ can be 
replaced by its angle average, using the following formula
\begin{eqnarray}
\frac{1}{(\mathbf{q}-\mathbf{q}')^2+\kappa^2}&\rightarrow&
\frac{\int_0^{\pi}\int_0^{2\pi}\frac{1}{(\mathbf{q}-\mathbf{q}')^2+\kappa^2}\sin{\theta}d\theta{d}\phi}{\int_{0}^{\pi}\int_0^{2\pi}\sin{\theta}d\theta{d}\phi}\nonumber\\
&=&\frac{1}{2}\int_0^{\pi}\frac{1}{(\mathbf{q}-\mathbf{q}')^2+\kappa^2}\sin{\theta}d\theta\nonumber\\
&=&
\frac{1}{4 q q'}\mathrm{log}[\frac{q^2+q'^2+2qq'+\kappa^2}{q^2+q'^2-2qq'+\kappa^2}].
\end{eqnarray}
The final expression depends only on the magnitudes of vectors $\mathbf{q}$ and $\mathbf{q}'$. 
The case of intranodal interactions is treated in the same way. 

\section{Derivation of the formulas for optical conductivity}\label{appendixE}
\subsection{Normal state: 2D and 3D DM}\label{appE_normalstate}
\subsubsection{Equilibrium}\label{appE_normalstate_eq}
The general expression for the real part of the frequency-dependent conductivity was introduced in Section~\ref{Opt_cond} of the main text [see Eq.~(\ref{eq:Kubo})]
%
%
In order to calculate the optical conductivity using Eq.~(\ref{eq:Kubo}), one needs to know the spectral function. 

We will now consider the non-interacting Weyl Hamiltonian for one node with a given chirality ($\xi=+1$) [see Eq.~(\ref{eq:Weyl_diag})]. 
The Green's function of the system in the electron/hole (diagonal) basis can be  
readily obtained by inverting the Hamiltonian
\begin{eqnarray}
 \tilde{G}(\mathbf{k},\omega)=\left(
 \begin{array}{lc}
  \frac{1}{\omega-\varepsilon_{\mathbf{k}}^{+}}  & 0 \\
  0  & \frac{1}{\omega-\varepsilon_{\mathbf{k}}^{-}} \\
 \end{array}
 \right),
\end{eqnarray}
where $\varepsilon_{\mathbf{k}}^{\pm}=\pm\hbar{v}|\mathbf{k}|$ are conduction/valence band dispersions. 

By performing the unitary transformation defined in Eq.~(\ref{eq:Weyl_unitary_transform}), we obtain the Green's function in the original spin basis
\begin{eqnarray}
  G(\mathbf{k},\omega)=\left(
 \begin{array}{lc}
  \frac{\omega+k_z}{\omega^2-(\hbar{v}|\mathbf{k}|)^2}  & \frac{k_x-i k_y}{\omega^2-(\hbar{v}|\mathbf{k}|)^2} \\
  \frac{k_x+i k_y}{\omega^2-(\hbar{v}|\mathbf{k}|)^2}  & \frac{\omega-k_z}{\omega^2-(\hbar{v}|\mathbf{k}|)^2} \\
 \end{array},
 \right)
\end{eqnarray}
The elements of the Green's function can be conveniently re-written as  
$G_{11}(\omega,\mathbf{k})=\frac{|\mathbf{k}|+k_z}{2|\mathbf{k}|(\omega-\hbar{v}|\mathbf{k}|)}+\frac{|\mathbf{k}|-k_z}{2|\mathbf{k}|(\omega-\hbar{v}|\mathbf{k}|)}$

From the Green's function, we obtain the spectral function in the diagonal basis
\begin{eqnarray}
 \tilde{A}(\mathbf{k},\omega)=\left(
 \begin{array}{lc}
  \delta(\omega-\varepsilon_{\mathbf{k}}^{+})  & 0 \\
  0  & \delta(\omega-\varepsilon_{\mathbf{k}}^{-}) \\
 \end{array}
 \right),
\end{eqnarray}
and in the spin basis
\begin{eqnarray}
 A(\mathbf{k},\omega)=\left(
 \begin{array}{lc}
  A_{11}(\mathbf{k},\omega)  & A_{12}(\mathbf{k},\omega) \\
  A_{21}(\mathbf{k},\omega)  & A_{22}(\mathbf{k},\omega) \\
 \end{array}
 \right),
\end{eqnarray}
with the following matrix elements: 
\begin{eqnarray}\label{eq:Weyl_spec}
 A_{11}(\mathbf{k},\omega) & = & u_k^2\delta(\omega-\varepsilon_{\mathbf{k}}^{+})+v_k^2\delta(\omega-\varepsilon_{\mathbf{k}}^{-}),\\
 A_{22}(\mathbf{k},\omega) & = & v_k^2\delta(\omega-\varepsilon_{\mathbf{k}}^{+})+u_k^2\delta(\omega-\varepsilon_{\mathbf{k}}^{-}),\\
 A_{12}(\mathbf{k},\omega) & = & w_k[\delta(\omega-\varepsilon_{\mathbf{k}}^{+})-\delta(\omega-\varepsilon_{\mathbf{k}}^{-})],\\
 A_{21}(\mathbf{k},\omega) & = & w_k^*[\delta(\omega-\varepsilon_{\mathbf{k}}^{+})-\delta(\omega-\varepsilon_{\mathbf{k}}^{-})].
\end{eqnarray}
The spectral weights are given by
\begin{eqnarray}\label{eq:Weyl_spec2}
 u_k^2&=&\frac{1}{2}(1+\frac{k_z}{|\mathbf{k}|}),\\
 v_k^2&=&\frac{1}{2}(1-\frac{k_z}{|\mathbf{k}|}),\\
 w_k&=&\frac{1}{|\mathbf{k}|}(k_x-ik_y).
\end{eqnarray}

We notice that $\hat{v}=\frac{\partial H}{\partial \mathbf{k}}=v\sigma$. In Eq.~(\ref{eq:Kubo}), the trace over velocities and spectral functions gives an expression proportional to 
$A_{11}(\mathbf{k},\omega)A_{22}(\mathbf{k},\omega+\Omega)+A_{22}(\mathbf{k},\omega)A_{11}(\mathbf{k},\omega+\Omega)+A_{12}(\mathbf{k},\omega)A_{12}(\mathbf{k},\omega+\Omega)+
A_{21}(\mathbf{k},\omega)A_{21}(\mathbf{k},\omega+\Omega)$. 
Next, we use the expressions for the matrix elements of $A$ [Eq.~(\ref{eq:Weyl_spec})-(\ref{eq:Weyl_spec2})]. The sum over momentum in the formula for optical conductivity  
[Eq.~(\ref{eq:Kubo})] can be 
transformed into an integral as $\sum_{\mathbf{k}}\rightarrow\int{d}^3k/(2\pi)^3$. The integral over 
momentum $\mathbf{k}$ can then be transformed into an integral over energy $\varepsilon$ after performing angular integration. 
Only the terms involving the diagonal elements of the spectral function remain after angular integration. 
Finally, we have
\begin{widetext}
\begin{eqnarray}
\mathrm{Re}[\sigma_{xx}(\Omega)] & = &\frac{e^2}{6\pi v_{F}\Omega}\int_{-\infty}^{+\infty}d\omega[n_{F}(\omega)-n_{F}(\omega+\Omega)]
 \times\int_{0}^{\infty}d\varepsilon\varepsilon^2[\delta(\omega-\varepsilon_{\mathbf{k}}^{+})\delta(\omega+\Omega-\varepsilon_{\mathbf{k}}^{+})+
\delta(\omega-\varepsilon_{\mathbf{k}}^{-})\delta(\omega+\Omega-\varepsilon_{\mathbf{k}}^{-})\nonumber\\
&\phantom{=}&\phantom{\frac{e^2}{6\pi v_{F}\Omega}\int_{-\infty}^{+\infty}d\omega[n_{F}(\omega)-n_{F}(\omega+\Omega)]
 \times} 
 +2(\delta(\omega-\varepsilon_{\mathbf{k}}^{+})\delta(\omega+\Omega-\varepsilon_{\mathbf{k}}^{-})+
\delta(\omega-\varepsilon_{\mathbf{k}}^{-})\delta(\omega+\Omega-\varepsilon_{\mathbf{k}}^{+}))].
\end{eqnarray}
\end{widetext}
There first two terms under the energy integral give the intraband  contribution (direct transitions within the same band) while  
the last two terms give the interband contribution (direct transitions between the two bands).

At $T=0$, one can perform integration analytically. This yields the following expression for the intraband (Drude) and interband piece, respectively
\begin{eqnarray}
\mathrm{Re}[\sigma_{xx}^\mathrm{intra}(\Omega)] & = & \frac{e^2\mu^2}{6\pi v_{F}}\delta(\Omega)\\
\mathrm{Re}[\sigma_{xx}^\mathrm{inter}(\Omega)] & = & \frac{e^2\Omega}{24\pi v_{F}}\Theta(\Omega-2\mu),
\end{eqnarray}
where $\mu>0$ is the chemical potential of the system and $\Theta(\omega)$ is a step function.

\subsubsection{Optical pumping}\label{appE_normalstate_pump}
In the case of optical pumping with population inversion~\cite{Svintsov_ape2014, Svintsov_optexpress2014}, 
one needs to take into account that the conduction and valence band are described by two distinct 
Fermi-Dirac distributions, 
$n^{e/h}_{\mathrm{F}}(\omega)=1/(e^{(\omega-\mu_{e/h})/T}+1)$, where 
 $n^{e(h)}_\mathrm{F}(\omega)$ is defined for $\omega>0$ ($\omega<0$). 
(In general, the electron and hole populations 
can also have different temperatures; however, for simplicity we will assume that after the population inversion has been established,  
 the photoexcited carriers can be described by a 
single electronic temperature $T$.) 

The calculation of the optical conductivity proceeds in the same way as for equilibrium but with $n^{e/h}_\mathrm{F}(\omega)$ instead of 
$n_{\mathrm{F}}(\omega)$. For pumped graphene, we have 
\begin{equation}
\mathrm{Re}[\sigma_{\mathrm{2D},xx}^\mathrm{inter}(\Omega),T]=\frac{e^2}{4\hbar}[n^{v}_{\mathrm{F}}(-\Omega/2)-n^{c}_{\mathrm{F}}(\Omega/2)].
\end{equation}
For $\mu_e=-\mu_h=\bar{\mu}$ (pumping in undoped graphene) and $T=0$, the conductivity becomes
\begin{equation}
\mathrm{Re}[\sigma_{xx}^\mathrm{inter}(\Omega),T=0] =\frac{e^2}{4\hbar}\mathrm{sgn}(\Omega-2\bar{\mu}),
\end{equation}
where $\mathrm{sgn}(\omega)$ is a sign function. 
One immediately notices that for $\Omega>2\bar{\mu}$, 
the conductivity is positive corresponding to direct interband transitions from valence band (occupied states below $\mu_h$) to conduction band 
(empty states below $\mu_e$) [see Fig.~\ref{opt_cond_scheme}]. If $\Omega<2\bar{\mu}$, the interband conductivity is negative.

In the case of a 3D DM, we get
\begin{equation}
\mathrm{Re}[\sigma_{\mathrm{3D},xx}^\mathrm{inter}(\Omega),T]=\frac{e^2}{24\pi{v}}[n^{v}_{\mathrm{F}}(-\Omega/2)-n^{c}_{\mathrm{F}}(\Omega/2)].
\end{equation}

\subsection{Ordered state}\label{appE_order}
Optical conductivity in the superconducting or excitonic state reads
\begin{align}\label{eq:opt_cond_SC_gen}
\mathrm{Re}[\sigma_{\alpha\beta}^\mathrm{SC/EX}(\Omega)] & =\frac{e^2\pi}{\Omega}\int_{-\infty}^{+\infty}d\omega[n_{F}(\omega)-n_{F}(\omega+\Omega)]\nonumber\\
& \times\sum_{\mathbf{k}}\mathrm{Tr}[\hat{v}_{\alpha}\mathcal{A}(\mathbf{k},\omega)\hat{v}_{\beta}\mathcal{A}(\mathbf{k},\omega+\Omega)],
\end{align}
where $\mathcal{A}$ is the full spectral function of the system
\begin{eqnarray}
 \mathcal{A}(\mathbf{k},\omega)=\left(
 \begin{array}{lc}
  A_1(\mathbf{k},\omega)  & B(\mathbf{k},\omega) \\
  B^{\dagger}(\mathbf{k},\omega)  & A_2(\mathbf{k},\omega) \\
 \end{array}
 \right),
\end{eqnarray}
Here $A_{1/2}(\mathbf{k},\omega)$ is the quasiparticle spectral function and 
 $B(\mathbf{k},\omega)$ is the anomalous spectral function. 
 
\subsubsection{Superconductor}\label{appE_order_bcs}
We will start by calculating the optical conductivity in the superconducting state~\cite{carbotte_prb2017_SC_opt}. 
A standard Hamiltonian of a superconductor reads
\begin{equation}
H = \sum_{\mathbf{k},\sigma}\varepsilon_{\mathbf{k}}c^{\dagger\sigma}_{\mathbf{k}}c_{\mathbf{k}}^{\sigma}+
\sum_{\mathbf{q},\mathbf{k},\mathbf{k}'}\sum_{\sigma,\sigma'}V(q)c^{\dagger\sigma}_{\mathbf{k+q}}c^{\dagger\sigma'}_{\mathbf{k'-q}}
c_{\mathbf{k}'}^{\sigma'}c_{\mathbf{k}}^{\sigma},
\end{equation}
where $c_{\mathbf{k}}^{\sigma}$ creates an electron with momentum $\mathbf{k}$ and spin $\sigma$ in the band with dispersion $\varepsilon_{\mathbf{k}}$. 
The last term is an effective attractive potential, $V(q)=-V_0$ for $|\varepsilon_{\mathbf{q}}|\le\hbar\omega_\mathrm{D}$ and zero otherwise, where $\omega_\mathrm{D}$ is
 the Debye frequency. We assume a simple quadratic dispersion ($\varepsilon_{\mathbf{k}}\propto k^2$).

The full Green's function $\mathcal{G}$ can be written as a matrix in the Nambu space
\begin{eqnarray}
 \mathcal{G}=\left(
 \begin{array}{lc}
  G^{\uparrow}(\mathbf{k},\tau-\tau')  & F(\mathbf{k},\tau-\tau') \\
  F^{\dagger}(\mathbf{k},\tau-\tau') & -G^{\downarrow}(-\mathbf{k},\tau'-\tau) 
 \end{array}
 \right).
\end{eqnarray}
Here $G^{\sigma}$, $\sigma=\uparrow,\downarrow$ is the single-particle Green's function and $F$ is the anomalous Green's function, which are defined as
\begin{eqnarray}
G^{\sigma}(\mathbf{k},\tau-\tau') & = & -<T_{\tau} c_{\mathbf{k}}^{\sigma}(\tau)c^{\dagger\sigma}_{\mathbf{k}}(\tau')>,\\
F(\mathbf{k},\tau-\tau') & = & -<T_{\tau}c_{\mathbf{k}}^{\uparrow}(\tau)c^{\downarrow}_{-\mathbf{k}}(\tau')>,\\
F^{\dagger}(\mathbf{k},\tau-\tau') & = & -<T_{\tau} c^{\dagger\downarrow}_{-\mathbf{k}}(\tau)c^{\dagger\uparrow}_{\mathbf{k}}(\tau')>,
\end{eqnarray}
where $T_{\tau}$ is the imaginary-time ordering operator. 

The mean-field superconducting order parameter, or gap, is defined as $\Delta_{\mathbf{k}}=T\sum_{\textbf{k}',i\omega_n}V_{\textbf{k}-\textbf{k}'}F(\textbf{k}';i\omega_n)$, 
where $\omega_n=(2n+1)\pi{T}$ are the Matsubara frequencies and $T$ is the electronic temperature. Since $V(\mathbf{q})$ is constant, the order 
parameter is also constant, $\Delta_{\mathbf{k}}=\Delta$. After solving the equations of motion for the Green's functions, we have
\begin{eqnarray}
G^{\sigma}(\mathbf{k},i\omega_n) & = & -\frac{i\omega_n+\varepsilon_{\mathbf{k}}}{\omega_n^2+\varepsilon_{\mathbf{k}}^2+\Delta^2},\\
F^{\dagger}(\mathbf{k},i\omega_n) & = & F(\mathbf{k},i\omega_n)=-\frac{\Delta}{\omega_n^2+\varepsilon_{\mathbf{k}}^2+\Delta^2}.
\end{eqnarray}
The poles of the Green's functions give the renormalized dispersions $\pm E_\mathrm{k}$, where
\begin{equation}
E_{\mathbf{k}}=\sqrt{\varepsilon_{\mathbf{k}^2}+\Delta^2}.
\end{equation}

Now the Green's function can be re-written as
\begin{eqnarray}
G^{\sigma}(\mathbf{k},i\omega_n) & = & \frac{u_{\mathbf{k}}^2}{i\omega_n-E_{\mathbf{k}}}+\frac{v_{\mathbf{k}}^2}{i\omega_n+E_{\mathbf{k}}},\\
F^{\dagger}(\mathbf{k},i\omega_n) & = & -u_{\mathbf{k}}v_{\mathbf{k}}(\frac{1}{i\omega_n-E_{\mathbf{k}}}-\frac{1}{i\omega_n+E_{\mathbf{k}}}),
\end{eqnarray}
where
\begin{eqnarray}
u_{\mathbf{k}}^2 & = & \frac{1}{2}(1+\frac{\varepsilon_{\mathbf{k}}}{E_{\mathbf{k}}}),\\
v_{\mathbf{k}}^2 & = & \frac{1}{2}(1-\frac{\varepsilon_{\mathbf{k}}}{E_{\mathbf{k}}}),
\end{eqnarray}

\begin{figure}[ht!]
\centering
\includegraphics[width=0.9\linewidth,clip=true]{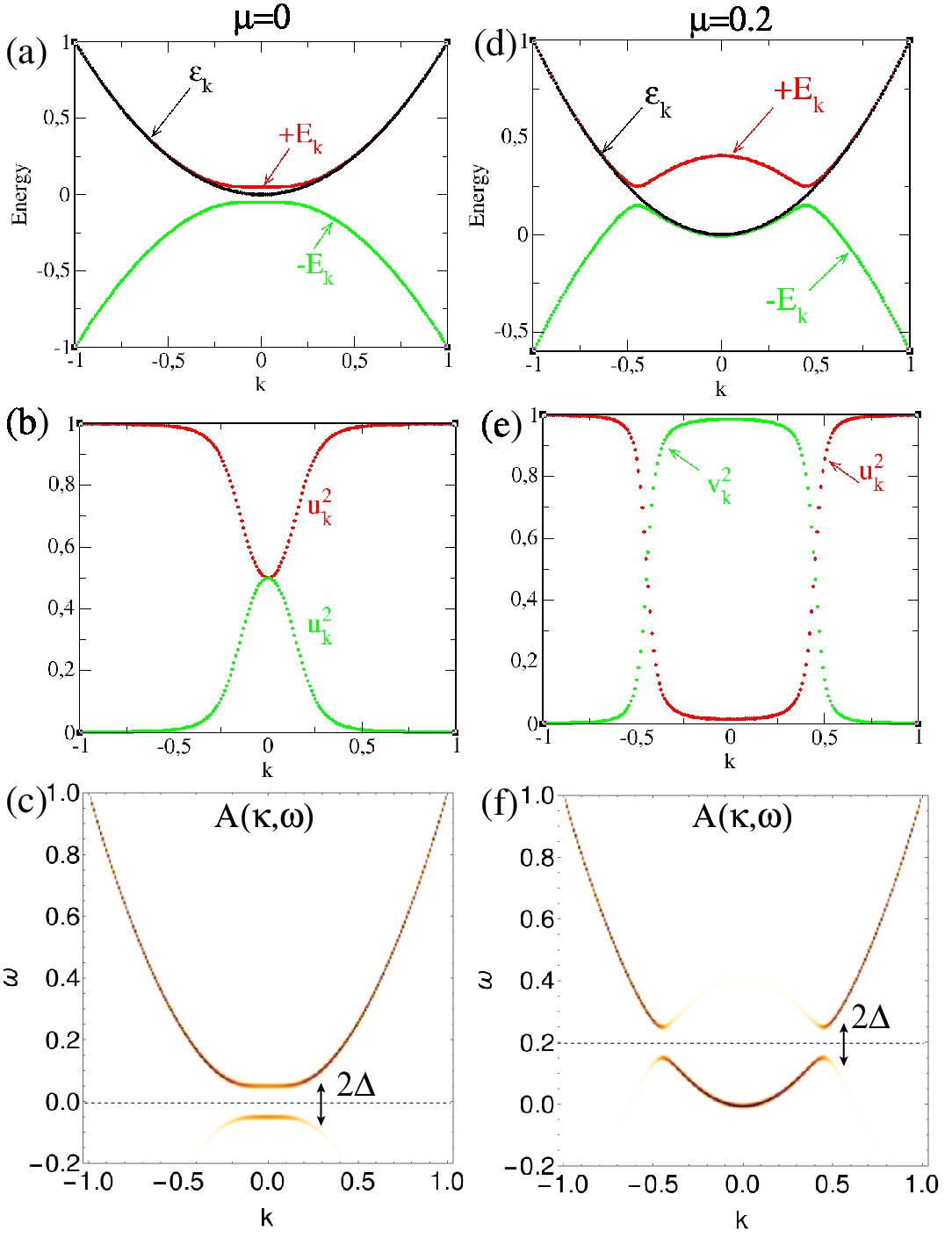}
\caption{Superconducting state. (a,d) Original and renormalized dispersions, (b,e) spectral weights, and (c,f) spectral function. Left: $\mu=0$, right: $\mu=0.2$. 
$\Delta$ is the superconducting gap.}
\label{fig:bcs}
\end{figure}

The quasiparticle and anomalous spectral functions are given by respectively
\begin{eqnarray}
A(\mathbf{k},\omega) & = & u_{\mathbf{k}}^2\delta(\omega-E_{\mathbf{k}})+v_{\mathbf{k}}^2\delta(\omega+E_{\mathbf{k}}),\label{eq:BCS_A}\\
B(\mathbf{k},\omega) & = & -u_{\mathbf{k}}v_{\mathbf{k}}(\delta(\omega-E_{\mathbf{k}})-\delta(\omega+E_{\mathbf{k}})).\label{eq:BCS_B}
\end{eqnarray}
The plots of the dispersions, the spectral weights and the spectral function are shown in Fig.~\ref{fig:bcs}.

The dynamical optical conductivity in the superconducting state becomes
\begin{eqnarray}\label{eq:cond_BCS}
\mathrm{Re}[\sigma^\mathrm{SC}(\Omega)] & = & \frac{e^2\pi}{\Omega}\int_{-\infty}^{+\infty}d\omega[n_{F}(\omega)-n_{F}(\omega+\Omega)]\nonumber\\
&\phantom{=}&\times \sum_{\mathrm{k}}v_\mathrm{F}^2(\mathrm{k})[A(\mathbf{k},\omega)A(\mathbf{k},\omega+\Omega)\nonumber\\
&\phantom{=}&\phantom{\times \sum_{\mathrm{k}}} +B(\mathbf{k},\omega)B(\mathbf{k},\omega+\Omega)],
\end{eqnarray}
where $v_\mathrm{F}$ is the Fermi velocity. Let us focus on the contribution coming from quasiparticle spectral function $A(\mathbf{k},\omega)$. 
Substituting Eq.~(\ref{eq:BCS_A}) into Eq.~(\ref{eq:cond_BCS}), for $T=0$
 and $\mu=0$, we obtain the intraband and interband contributions to the optical conductivity
\begin{eqnarray}\label{eq:cond_BCS1}
\mathrm{Re}[\sigma^\mathrm{SC}_\mathrm{intra}(\Omega)] & = &\frac{\mathcal{C}}{\Omega}\delta(\Omega)\int_{-\infty}^{+\infty}d\varepsilon\int_{-\Omega}^{0}d\omega
[u_{\mathbf{k}}^2\delta(\varepsilon-E_{\mathbf{k}})\nonumber\\
&\phantom&\phantom{\frac{\mathcal{C}}{\Omega}\delta(\Omega)\int_{-\infty}^{+\infty}d\varepsilon}
+v_{\mathbf{k}}^2\delta(\varepsilon+E_{\mathbf{k}})],\\
\mathrm{Re}[\sigma^\mathrm{SC}_\mathrm{inter}(\Omega)] & = & \frac{\mathcal{C}}{\Omega}\int_{-\infty}^{+\infty}d\varepsilon\int_{-\Omega}^{0}d\omega\delta(2\omega+\Omega)
[\delta(\varepsilon-E_{\mathbf{k}})\nonumber\\
&\phantom{=}&\phantom{\frac{\mathcal{C}}{\Omega}\int_{-\infty}^{+\infty}d\varepsilon\int_{-\Omega}^{0}d\omega
}+\delta(\varepsilon+E_{\mathbf{k}})],
\end{eqnarray}
where $\mathcal{C}=e^2 v_F^2 N(0)/2$ and $N(0)$ is the density of states in the normal state at $\mu=0$.

\begin{figure}[ht!]
\centering
\includegraphics[width=0.9\linewidth,clip=true]{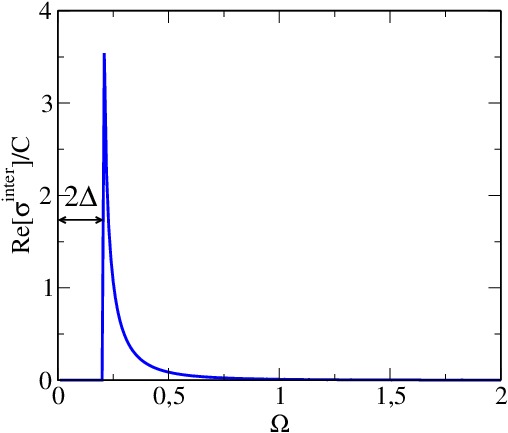}
\caption{Interband optical conductivity in the superconducting state for $T=0$ and $\mu=0$. $\Delta$ is the superconducting gap 
and 2$\Delta$ is the optical gap.}
\label{fig:bcs_opt_cond}
\end{figure}

The intraband conductivity is $0$, while the interband conductivity is:
\begin{eqnarray}\label{eq:cond_BCS_inter}
\mathrm{Re}[\sigma^\mathrm{SC}_\mathrm{inter}(\Omega)] = \Bigg\lbrace
\begin{array}{l}
\frac{\Lambda}{\Omega^2}\frac{\Delta^2}{\sqrt{\Omega^-(2\Delta)^2}},\quad \Omega>2\Delta \\ 
0,\quad \Omega<2\Delta.
\end{array}
\end{eqnarray}
The interband optical conductivity in the superconducting state (the quasiparticle contribution) is plotted in Fig.~(\ref{fig:bcs_opt_cond})

\subsubsection{Excitonic insulator}\label{appE_order_ex}
\begin{figure}[ht!]
\centering
\includegraphics[width=0.9\linewidth,clip=true]{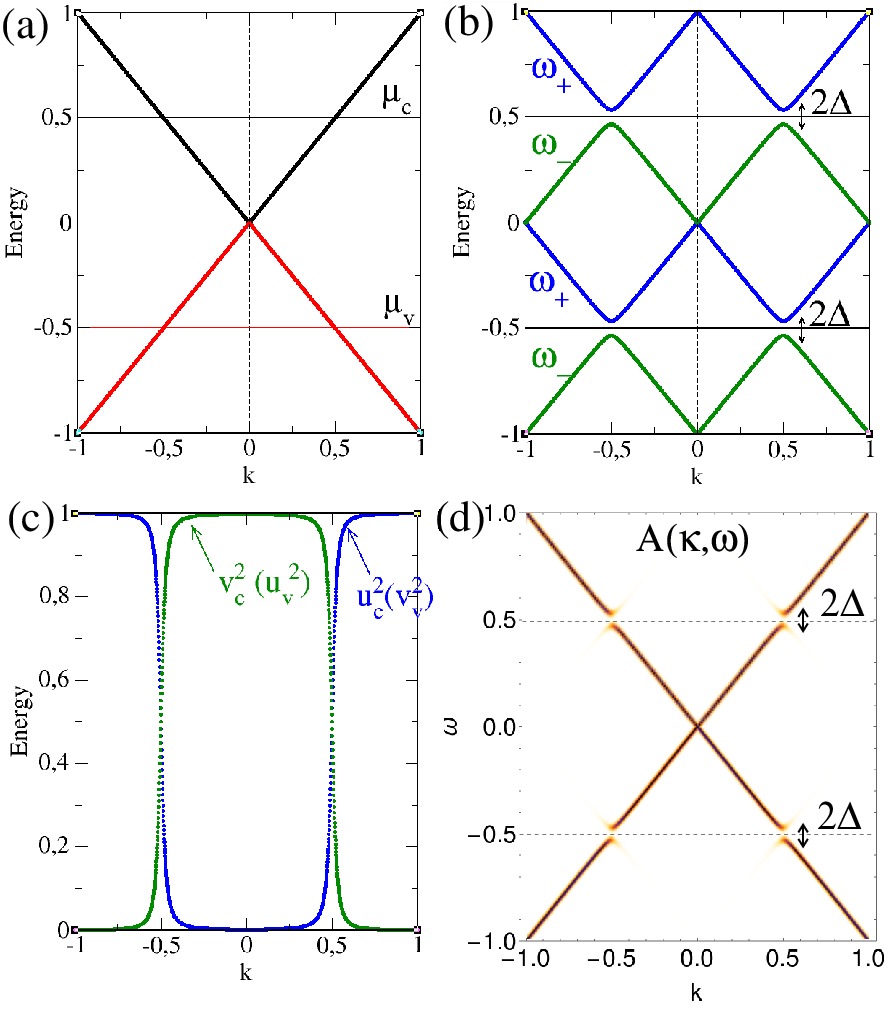}
\caption{Excitonic insulator. (a) Conduction and valence band dispersions, (b) renormalized dispersions, (c) spectral weights, (d) spectral function. $\Delta$ is the excitonic gap.}
\label{fig:ei}
\end{figure}

In the case of an excitonic insulator~\cite{halperin1968possible, jerome1967excitonic}, we have conduction and valence bands $\varepsilon_{\mathbf{k}}^{\pm}$. For a pumped 
DM these are given in Eqs.~(\ref{bands_pumped_CB}) and (\ref{bands_pumped_VB}).  
The full Green's function $\mathcal{G}$ of the system can be written as
\begin{eqnarray}
 \mathcal{G}=\left(
 \begin{array}{lc}
  G_{e}  & F \\
  F^{\dagger} & G_{h} 
 \end{array}
 \right),
\end{eqnarray}
where $G_e$ and $G_h$ are the single-particle Green's functions for electrons and holes, respectively, and $F$ is the anomalous Green's function, which are defined as
\begin{eqnarray}
G_e(\mathbf{k},\tau-\tau') & = & -<T_{\tau} c_{\mathbf{k},+}(\tau)c^{\dagger}_{\mathbf{k},+}(\tau')>,\\
G_h(\mathbf{k},\tau-\tau') & = & -<T_{\tau} c_{\mathbf{k},-}(\tau)c^{\dagger}_{\mathbf{k},-}(\tau')>,\\
F(\mathbf{k},\tau-\tau') & = & -<T_{\tau} c_{\mathbf{k},+}(\tau){c_{\mathbf{k},-}}^{\dagger}(\tau')>,\\
F^{\dagger}(\mathbf{k},\tau-\tau') & = & -<T_{\tau} c^{\dagger}_{\mathbf{k},-}(\tau)c_{\mathbf{k},+}(\tau')>.
\end{eqnarray}
Here $c^{\dagger}_{\mathbf{k},\pm}(c_{\mathbf{k},\pm})$ are the fermionic creation (annihilation) operators corresponding to bands $\varepsilon^{\pm}_{\mathbf{k}}$. 
The order parameter, or gap, for the excitonic condensate is defined as $\Delta_{\textbf{k}}=T\sum_{\textbf{k}',i\omega_n}V_{\textbf{k}-\textbf{k}'}F(\textbf{k}';i\omega_n)$. 
After solving the equations of motion for the Green's functions, we get
\begin{eqnarray}
G_e(\mathbf{k},\omega) & = & \frac{\omega-\varepsilon^{-}_{\mathbf{k}}}{(\omega-\varepsilon^{+}_{\mathbf{k}})(\omega-\varepsilon^{-}_{\mathbf{k}})-|\Delta_{\mathbf{k}}|^2},\\
G_h(\mathbf{k},\omega) & = & \frac{\omega-\varepsilon^{+}_{\mathbf{k}}}{(\omega-\varepsilon^{+}_{\mathbf{k}})(\omega-\varepsilon^{-}_{\mathbf{k}})-|\Delta_{\mathbf{k}}|^2},\\
F^{\dagger}(\mathbf{k},\omega_) & = & -\frac{\Delta_{\mathbf{k}}}{(\omega-\varepsilon^{+}_{\mathbf{k}})(\omega-\varepsilon^{-}_{\mathbf{k}})-|\Delta_{\mathbf{k}}|^2}.
\end{eqnarray}
The poles of the Green's functions give the renormalized dispersions [see also Eq.~(\ref{exc_bands})]
\begin{equation}
 \omega_{\pm}(\mathbf{k})=\frac{\varepsilon^{+}_{\mathbf{k}}+\varepsilon^{-}_{\mathbf{k}}}{2}\pm\frac{1}{2}
 \sqrt{(\varepsilon^{+}_{\mathbf{k}}-\varepsilon^{-}_{\mathbf{k}})^2+4|\Delta_(\mathbf{k})|^2}.
\end{equation}
Note that in our derivation for the excitonic insulator we have shifted the conduction band energies by $-\mu_e$ ($\mu_e>0$) and 
the valence band energies by $\mu_h$ ($\mu_h<0$). Therefore when plotting the quantities associated with conduction or valence band, we have to shift back the energy 
in order to restore the original positions of the two chemical potentials (see Fig.~\ref{fig:ei}). Note also that we have two copies of the excitonic bands, one for the conduction band and one for the valence 
band, with the corresponding spectral weights introduced below. 

The spectral functions of the excitonic insulator are given by
\begin{eqnarray}
A_e(\mathbf{k},\omega) & = & u_{e\mathbf{k}}^2\delta(\omega-\omega_{+})+v_{e\mathbf{k}}^2\delta(\omega-\omega_{-}),\label{eq:aE_EI_Ac}\\
A_h(\mathbf{k},\omega) & = & u_{h\mathbf{k}}^2\delta(\omega-\omega_{+})+v_{h\mathbf{k}}^2\delta(\omega-\omega_{-}),\label{eq:aE_EI_Av},\\
B(\mathbf{k},\omega) & =  & -\frac{\Delta_{\mathbf{k}}}{\omega_{+}-\omega_{-}}[\delta(\omega-\omega_{+})-\delta(\omega-\omega_{-}].\label{eq:aE_EI_B}
\end{eqnarray}
where the spectral weights are given by
\begin{eqnarray}
\begin{array}{lc}
u^2_{e\mathbf{k}}  = \frac{\omega_{+}-\varepsilon^{-}_{\mathbf{k}}}{\omega_{+}-\omega_{-}}, & v^2_{e\mathbf{k}} = \frac{\omega_{-}-\varepsilon^{-}_{\mathbf{k}}}{\omega_{-}-\omega_{+}},\\
u^2_{h\mathbf{k}}  = \frac{\omega_{+}-\varepsilon^{+}_{\mathbf{k}}}{\omega_{+}-\omega_{-}}, & v^2_{h\mathbf{k}} = \frac{\omega_{-}-\varepsilon^{+}_{\mathbf{k}}}{\omega_{-}-\omega_{+}}.
\end{array}
\end{eqnarray}
The plots of the dispersions, the spectral weights and the spectral function are shown in Fig.~\ref{fig:bcs}.
Substituting the expressions for the spectral functions [Eqs.~(\ref{eq:aE_EI_Ac})-(\ref{eq:aE_EI_B})] into the equation  
for the optical conductivity [Eq.~(\ref{eq:opt_cond_SC_gen})], we obtain Eq.~(\ref{eq:opt_cond_pump}) of the main text.

\bibliography{Driven_Dirac}

\begin{thebibliography}{67}
\bibitem [{num({\natexlab{a}})}]{numerical_note}%
  \BibitemOpen
  \href@noop {} {}\bibinfo {howpublished} {For numerical calculations with the
  screened Coulomb potential in equilibrium ($\mu=0$), we use a small but
  finite constant value of the screening wavevector; variations of this value
  do not change the results appreciably.} ({\natexlab{a}})\BibitemShut
  {NoStop}%
\end{thebibliography}

\end{document}